\documentclass[prb, twocolumn, showpacs, superscriptaddress]{revtex4-1}
\usepackage{amsmath, amssymb}
\usepackage{graphicx}
\usepackage{setspace}
\usepackage{color}

\begin{document}
\title{Inelastic Kondo-Andreev tunnelings in a vibrating quantum dot}
%\date{\today}
\author{Zhan Cao}
\affiliation{Center for Interdisciplinary Studies and Key Laboratory for Magnetism and Magnetic Materials of the Ministry of Education, Lanzhou University, Lanzhou 730000, China}
\author{Tie-Feng Fang}
\email{fangtiefeng@lzu.edu.cn}
\affiliation{Center for Interdisciplinary Studies and Key Laboratory for Magnetism and Magnetic Materials of the Ministry of Education, Lanzhou University, Lanzhou 730000, China}
\author{Qing-Feng Sun}
\affiliation{International Center for Quantum Materials, School of Physics, Peking University, Beijing 100871, China}
\affiliation{Collaborative Innovation Center of Quantum Matter, Beijing 100871, China}
\author{Hong-Gang Luo}
\email{luohg@lzu.edu.cn}
\affiliation{Center for Interdisciplinary Studies and Key Laboratory for Magnetism and Magnetic Materials of the Ministry of Education, Lanzhou University, Lanzhou 730000, China}
\affiliation{Beijing Computational Science Research Center, Beijing 100084, China}
\begin{abstract}
Phonon-assisted electronic tunnelings through a vibrating quantum dot embedded between normal and superconducting leads are studied in the Kondo regime. In such a hybrid device, with the bias applied to the normal lead, we find a series of Kondo sidebands separated by half a phonon energy in the differential conductance, which are distinct from the phonon-assisted sidebands previously observed in the conventional Andreev tunnelings and in systems with only normal leads. These Kondo sidebands originate from the Kondo-Andreev cooperative cotunneling mediated by phonons, which exhibit a novel Kondo transport behavior due to the interplay of the Kondo effect, the Andreev tunnelings, and the mechanical vibrations. Our result could be observed in a recent experiment setup [J. Gramich \emph{et al.}, PRL \textbf{115}, 216801 (2015)], provided that their carbon nanotube device reaches the Kondo regime at low temperatures.
\end{abstract}

\maketitle
\emph{Introduction.---}The hybrid quantum systems have a potential to exhibit new emergent phenomena through merging the strength of different media \cite{Schleier2016}. A quantum dot (QD) embedded between normal (N) and s-wave superconducting (S) leads (N-QD-S) is one of such devices, which has received considerable attentions from both the theoretical \cite{Fazio1998,Sun2001,Clerk2000,Domanski2007,Tanaka2007,Yamada2011,Baranski2013,Koga2013,Zitko2015,Li2015,Domanski2016} and experimental \cite{Graber2004,Deacon2010a,Deacon2010b} communities in the past two decades. In such a hybrid system, two important phenomena may arise: one is the Andreev tunneling (AT) \cite{Andreev1964} and the other is the screening of the localized spin in the QD by conduction electrons in the leads. While the former induces the Andreev bound states (ABSs) located in the supreconducting gap, the latter is the famous Kondo effect \cite{Hewson1993}. The competition between these two processes results in a profound influence on the ground state properties \cite{Tanaka2007,Yamada2011,Baranski2013,Koga2013,Zitko2015,Li2015,Domanski2016} as well as the transport behaviors of the devices \cite{Fazio1998,Sun2001,Clerk2000,Domanski2007,Tanaka2007,Yamada2011,Baranski2013,Koga2013,Li2015,Domanski2016}.

For a molecular QD, it was found that vibrational degrees of freedom are easily excited when electronic tunneling takes place \cite{Reed1997,Park2000,Ventra2001}, which has a dramatic influence on the transport of the system due to the presence of inelastic tunneling processes mediated by emission or absorption of phonons \cite{LeRoy2004,Sapmaz2006,Huttel2009,Leturcq2009}. In recent years, phonon-assisted inelastic AT in an N-QD-S system also leads to interesting physics on, for example, the electronic transport \cite{Bai2011,Baranski2015b,Bocian2015}, the heat generation \cite{Wang2013}, the ground-state cooling \cite{Stadler2016}, the steady-state shot noise \cite{Zhang2009}, as well as the transient dynamics under a step bias \cite{Albrecht2013}. More interestingly, the phonon-assisted AT can lead to resonant peaks every time the bias voltage changes by one phonon energy or the gate voltage changes by half a phonon energy \cite{Zhang2012,Baranski2015a}, which has been unambiguously observed in a recent experiment \cite{Gramich2015}. This is somewhat reminiscent of normal systems where phonon sidebands of Kondo cotunnelings \cite{Konig1996,Chen2006,Fernandez2008,Roura2013,Rakhmilevitch2014,Roura2016} and single-electron tunnelings \cite{Chen2005,Galperin2006,Fang2011} are also separated by one phonon energy in the bias voltage. Since the N-QD-S setup fabricated in the experiment \cite{Gramich2015} is indeed an ideal platform to explore the Kondo physics, it is our aim in this paper to provide a theoretical study of the Kondo transport in such a device.

\begin{figure}[t]
\centering
\includegraphics[width=0.95\columnwidth]{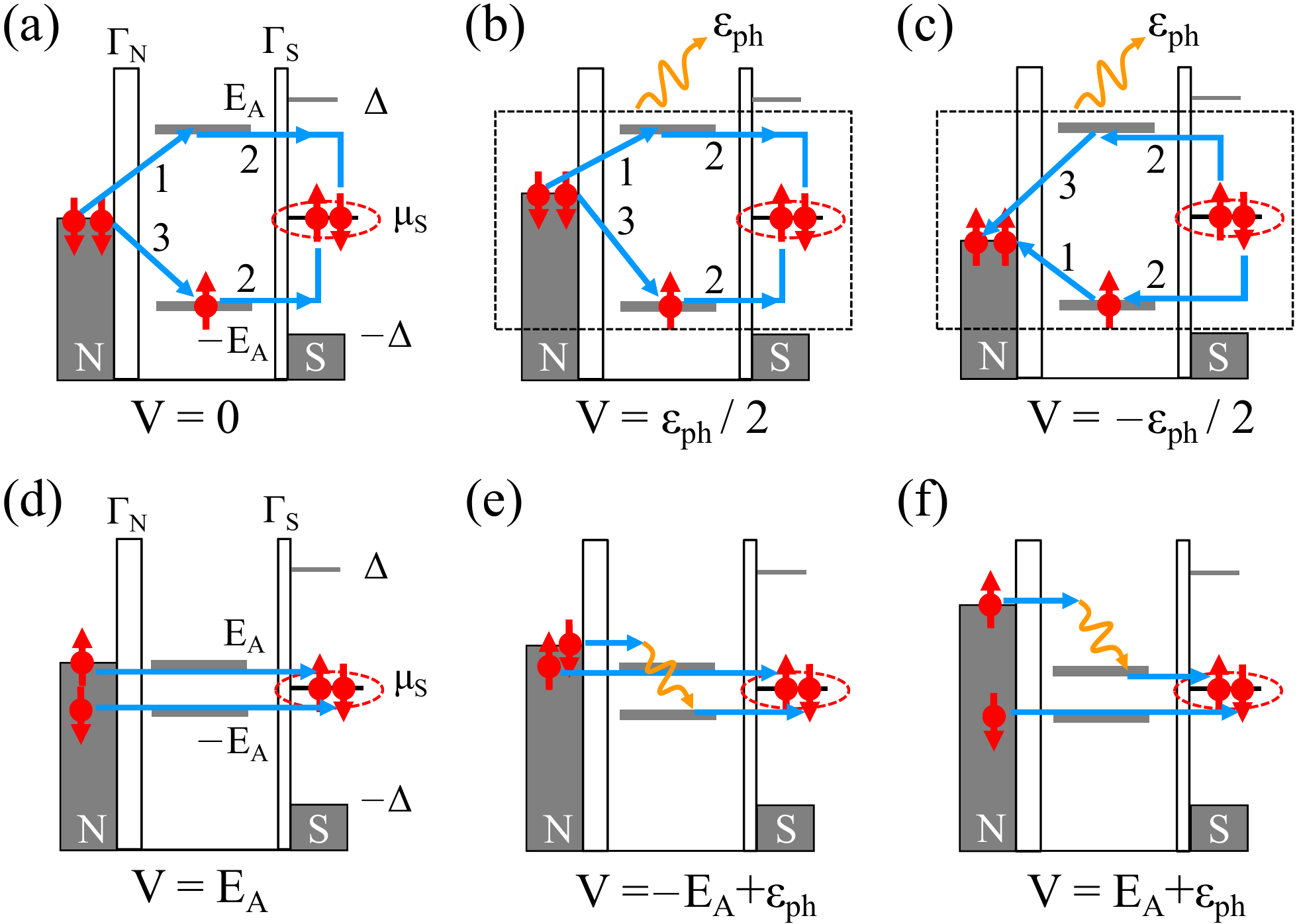}
\caption{(Color online) Schematics of elastic [(a), (d)] and inelastic [(b), (c), (e), (f)] electronic tunnelings in an N-QD-S system. (a)-(c) represent the Kondo-Andreev tunnelings through an interacting QD with the numbers $1,\,2,\,3$ denoting the tunneling sequences. (d)-(f) show the conventional AT through a noninteracting QD. The solid lines in the QD indicate the two ABSs with energies $\pm E_A$, while the wavy arrows represent the emission of phonons during the inelastic tunnelings.}\label{fig1}
\end{figure}

Our investigation reveals that the interplay of the Kondo correlations, the superconductivity, and the mechanical vibrations of the QD gives rise to distinct transport characteristics, as compared with those arising from the conventional phonon-assisted ATs \cite{Zhang2012,Baranski2015a,Gramich2015}. The main physical scenario is illustrated in Fig.\,\ref{fig1}, where elastic and inelastic AT with and without the Kondo effect are schematically shown. We set the chemical potentials of the N ($\mu_N$) and S ($\mu_S$) leads as $\mu_N=V$, $\mu_S=0$, and the superconducting gap $\Delta$ is taken as the largest energy scale in the problem. We consider the parameter regime where the QD-S tunnel coupling is much larger than the N-QD coupling and both are several times smaller than the onsite Coulomb repulsion, such that the Kondo effect and the onsite pairing coexist \cite{Deacon2010b}. In this case, dot electrons would undergo frequent Andreev reflections at the QD-S interface, which forms two Andreev bound states (ABSs) with energies $\pm E_A$ in the spectrum of QD. The ABSs are separated roughly by the Coulomb energy and their widths are determined by the N-QD coupling \cite{Baranski2013}. At zero bias $V=0$, a spin-$\uparrow$ localized electron and a spin-$\downarrow$ lead electron at $\mu_N$ can convert to a Cooper pair in S, while another spin-$\downarrow$ lead electron at $\mu_N$ transits into the QD simultaneously [Fig.\,\ref{fig1}(a)]. This spin-flip cotunneling process, to which we refer as the Kondo-Andreev tunneling, is elastic and accounts for the zero-bias conductance peak previously observed in this system \cite{Graber2004,Deacon2010b}. When the bias increases to $V=\varepsilon_{ph}/2$ ($\varepsilon_{ph}$ the phonon energy), besides the elastic Kondo-Andreev tunneling process, additional inelastic Kondo-Andreev tunneling emitting one phonon can also take place [Fig.\,\ref{fig1}(b)]. Here, the emission of a phonon fulfills the energy conservation of the transition that two N-lead electrons each with energy $\varepsilon_{ph}/2$ in the initial state are annihilated and a Cooper pair with zero energy is created in the final state, while the QD energy under the spin flipping remains the same. At negative bias $V=-\varepsilon_{ph}/2$, similar inelastic Kondo-Andreev tunneling can occur from the S lead to the N lead [Fig.\,\ref{fig1}(c)]. The opening of these additional tunneling channels would give rise to additional conductance peaks at $V=\pm\varepsilon_{ph}/2$. When multiple-phonon processes are involved, a series of sidebands separated by half a phonon energy are thus expected at $V=n\varepsilon_{ph}/2$ with $n=0,\pm1,\pm2\cdots$.

For comparison, we also give a general scenario of the conventional phonon-assisted ATs for a noninteracting N-QD-S, where the QD without the onsite Coulomb interaction favors even electron occupation and the distance between the two ABSs are determined roughly by the QD-S coupling \cite{Baranski2013}. In this system, there are two interleaved sets of phonon sidebands, each separated by $\varepsilon_{ph}$, in the differential conductance, since additional phonon-emitted inelastic AT can be triggered at $V=\pm E_A+n\varepsilon_{ph}$. For $n=0$, the AT is elastic [see Fig.\,1(d) for $V=E_A$]. For $n>0$, an N-lead electron at $\mu_N$ can transfer to the S lead through the lower [Fig.\,1(e), $V=-E_A+n\varepsilon_{ph}$] or upper [Fig.\,1(f), $V=E_A+n\varepsilon_{ph}$] ABS by emitting $n$ phonons, while another electron passes directly through the other ABS. Similar inelastic ATs take place from the S lead to the N lead for $n<0$. When the two ABSs are indistinguishable (e.g., their widths being larger than their interval) or separated by multiples of $\varepsilon_{ph}$, the two sets of phonon sidebands merge into a single set of sidebands separated by one phonon energy. This is exactly the special case discussed in Ref.\,\onlinecite{Zhang2012}. In the following, we perform a model calculation to demonstrate these transport scenarios.

\emph{Model and Formalism.---}Our N-QD-S system is modeled by the Hamiltonian $H=H_{\textrm{leads}}+H_{\textrm{ph}}+H_{\textrm{QD}}+H_{\textrm{tunnel}}$. The first term represents the normal ($\beta=\textrm{N}$) and superconducting ($\beta=\textrm{S}$) leads, $H_{\textrm{leads}}=\sum_{k,\sigma,\beta}\varepsilon_{k} c_{k\sigma\beta}^\dag c_{k\sigma\beta}-\Delta\sum_{k}(c_{k\uparrow S}^{\dag}c_{-k\downarrow S}^{\dag}+c_{-k\downarrow S}c_{k\uparrow S})$. $H_{\textrm{ph}}=\varepsilon_{ph} a^\dag a$ models the local phonon mode. $H_{\textrm{QD}}=\sum_{\sigma}\varepsilon_{d}d_{\sigma}^{\dag}d_{\sigma}+Un_{d\uparrow}n_{d\downarrow}+\lambda(a+a^\dag)\sum_\sigma n_{d\sigma}$ describes an interacting single-level QD, with Coulomb repulsion energy $U$, coupled with the local phonon by $\lambda$ the Holstein-type electron-phonon interaction (EPI). The last term $H_{\textrm{tunnel}}=\sum_{k,\sigma,\beta}(V_{\beta} c_{k\sigma\beta}^\dag d_{\sigma}+\textrm{H.c.})$ describes the electronic tunneling between the dot and the leads. From the tunneling matrix elements $V_{\beta}$, the dot level $\varepsilon_d$ acquires an intrinsic broadening $\Gamma_{\beta}\equiv 2\pi\rho_{_0}|V_\beta|^2$ with $\rho_{_0}$ the density of states of lead N and lead S in normal state. By the standard Keldysh nonequilibrium Green's function (GF) theory \cite{Haug2008}, the electronic current flowing from the N lead into the QD can be expressed as
\begin{eqnarray}
I=\frac{2ie}{h}\int \textrm{d}\omega\,\Gamma_{\textrm{N}}[(1-f_N)\mathbf{G}_{11}^{<}(\omega)+f_N\mathbf{G}_{11}^{>}(\omega)],\label{eq1}
\end{eqnarray}
where $f_{\textrm{N}}(\omega)$ is the Fermi distribution function of lead N. The boldfaced GF matrices are defined in the well-known $2\times2$ Nambu representation \cite{Sun2000}, from which the local density of states (LDOS) per spin can be calculated by $\rho(\omega)=-(1/\pi)\text{Im}\mathbf{G}^r_{11}(\omega)$.

Due to the presence of EPI, calculating the GFs needed in the current and the LDOS is nontrivial \cite{Hewson2002,Galperin2006}, even if the QD itself is noninteracting. Various approximations treating the EPI from the weak to strong coupling regime and from equilibrium to nonequilibrium have been established \cite{Cornaglia2004,Galperin2007a,Zimbovskaya2011}. In this work, we focus on the strong EPI regime. It is thus appropriate to make the non-perturbative Lang-Firsov transformation \cite{Lang1963} $\tilde{H}=e^{S}He^{-S}$ with $S=(\lambda/\varepsilon_{ph})(a^{\dag}-a)\sum_{\sigma}n_{d\sigma}$ to eliminate the linear EPI. This gives us $\tilde{H}=H_{\textrm{leads}}+H_{\textrm{ph}}+\tilde{H}_{\textrm{QD}}+\tilde{H}_\textrm{tunnel}$, where $\tilde{H}_{\textrm{QD}}=\sum_{\sigma}\tilde{\varepsilon }_{d}n_{d\sigma}+\tilde{U}n_{d\uparrow}n_{d\downarrow}$ and $\tilde{H}_\textrm{tunnel}=\sum_{k,\sigma,\beta}( \tilde V_\beta c_{k\sigma\beta}^\dag d_{\sigma}+\textrm{H.c.})$, with $\tilde{\varepsilon}_{d}=\varepsilon_{d}-g\varepsilon_{ph}$, $\tilde{U}=U-2g\varepsilon_{ph}$, $\tilde{V}_{\beta}=V_{\beta}X$, and $X=\textrm{exp}[-(\lambda/\varepsilon_{ph})(a^{\dag}-a)]$. Here a dimensionless measure of EPI $g\equiv\lambda^{2}/\varepsilon_{ph}^{2}$ is introduced. As in dealing with the localized polarons, we adopt the approximation replacing the operator $X$ with its expectation value $\langle X\rangle=\textrm{exp}[-g(N_\textrm{ph}+1/2)]$, where the average is taken over the independent phonon bath $H_\textrm{ph}$, and $N_\textrm{ph}$ is the Bose distribution. Hence, the renormalized $\tilde\Gamma_\beta=\langle X\rangle^2\Gamma_\beta$. This zero-order approximation which ignores the backaction of electrons on the phonons is valid when $V_\beta\ll\lambda$ and has been widely employed in the literature \cite{Kuo2002,Chen2005,Chen2006,Bai2011,Fang2011,Baranski2015a,Baranski2015b,Bocian2015,Wang2013,Zhang2009}. Previous studies \cite{Galperin2006} which compares a full self-consistent calculation and the zero-order approximation shows that the latter can predict accurate positions of the phonon sidebands, even though their exact lineshapes are missed to some extent. This suffices the purpose of our work. Applying the above decoupling scheme and the Feynman disentangling technique \cite{Mahan2000}, one obtains $\mathbf{G}_{11}^{r}(\omega)=\sum_{n=-\infty}^{\infty}L_{n}[\mathbf{\tilde{G}}_{11}^{r}(\omega-n\varepsilon_{ph})
+\frac{1}{2}\mathbf{\tilde{G}}_{11}^{<}(\omega-n\varepsilon_{ph})-\frac{1}{2}\mathbf{\tilde{G}}_{11}^{<}(\omega+ n\varepsilon_{ph})]$ and $\mathbf{G}_{11}^{<(>)}(\omega)=\sum_{n=-\infty}^{\infty}L_{n}\mathbf{\tilde{G}}_{11}^{<(>)}(\omega\pm n\varepsilon_{ph})$, where $L_{n}=\textrm{exp}[-g(2N_\textrm{ph}+1)]\textrm{exp}[n\beta\varepsilon_{ph}/2]I_{n}(x)$, with $x=2g\sqrt{N_\textrm{ph}(N_\textrm{ph}+1)}$ and $I_n(x)$ being the modified Bessel function of the first kind. Note that the new GFs $\tilde{\mathbf{G}}$ is defined according to the Hamiltonian $\tilde{H}$ in which the Bose degrees of freedom is totally decoupled.

We solve the retarded GF $\mathbf{\tilde{G}}^{r}$ using the equation-of-motion method \cite{Lacroix1981,Entin2005,Kashcheyevs2006}. This method forms probably one of the simplest basis for qualitatively capturing the Kondo physics and thus has been widely used in the literature \cite{Fazio1998,Sun2001,Domanski2007,Baranski2013,Li2015,Meir1991,Meir1993,Swirkowicz2003,Luo2004,Galperin2007,Zimbovskaya2008,Fang2008,Monreal2009,Lim2013,Krychowski2014,Xin2015,Hoffman2015}. Here, $\mathbf{\tilde{G}}^{r}$ is solved under the truncation scheme previously adopted by Sun \emph{et al}.\,\cite{Sun2001} (see details in the Supplemental Material \cite{SM}). The lesser and greater GFs $\mathbf{\tilde{G}}^{<(>)}$ are then obtained through the Keldysh equation $\mathbf{{\tilde G}}^{<(>)}=\mathbf{{\tilde G}}^{r}\mathbf{{\tilde \Sigma}}^{<(>)}\mathbf{{\tilde G}}^{a}$ with $\mathbf{\tilde G^a}=(\mathbf{\tilde G^r})^\dag$. Having these GFs self-consistently determined \cite{SM}, the current $I$, differential conductance $G\equiv dI/dV$, and LDOS $\rho(\omega)$ can be directly calculated. In the Supplemental Material \cite{SM}, we also use the modified second-order perturbation theory in the Coulomb interaction \cite{Yamada2011} to calculate $\rho(\omega)$ and $G$, which agrees with and complements the equation-of-motion results here.

\begin{figure}[t]
\centering
\includegraphics[width=0.95\columnwidth]{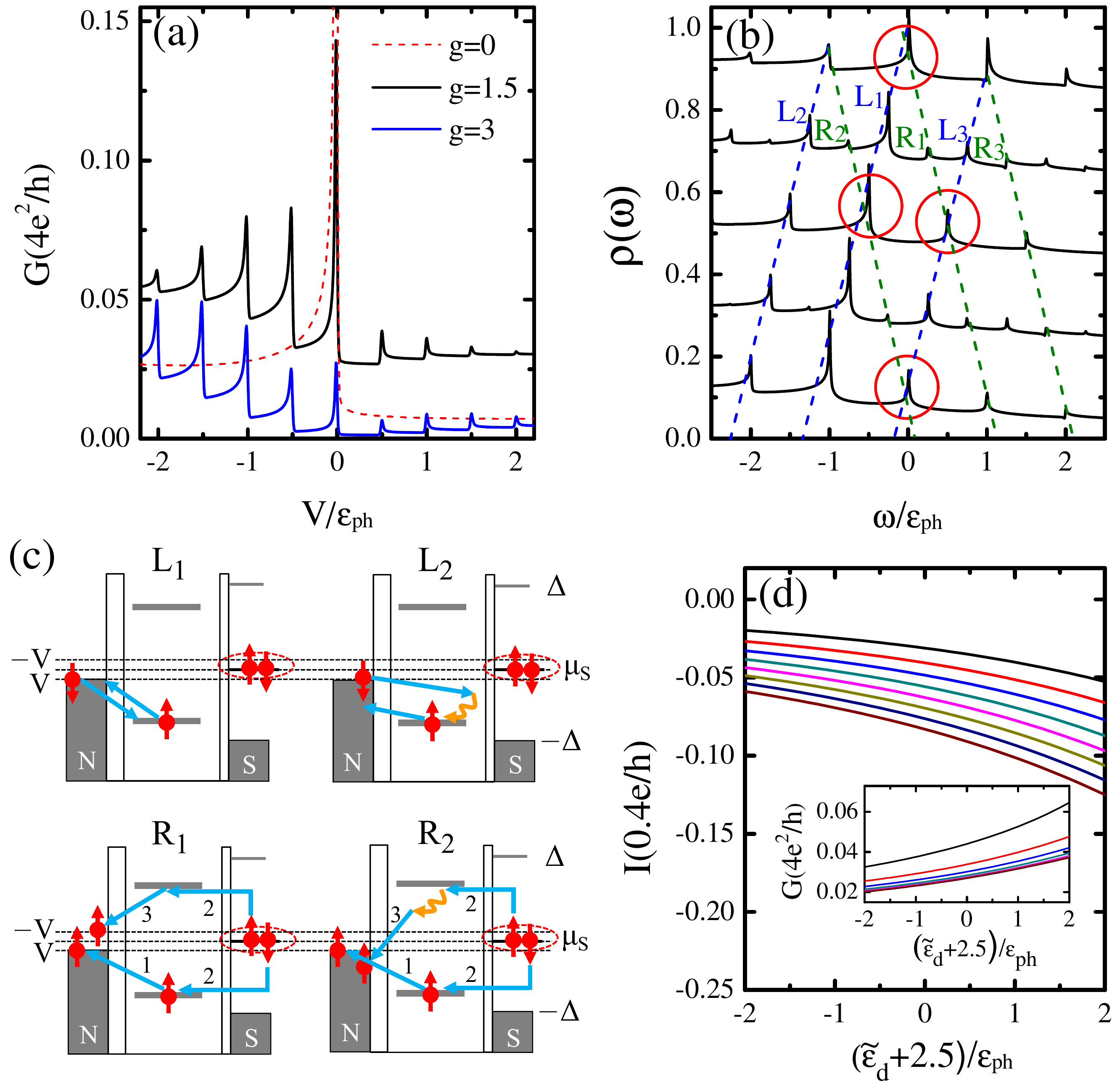}
\caption{(Color online) (a) Differential conductance in the Kondo regime with different EPI strength for $\tilde\varepsilon_d=-2.5$, $\tilde\Gamma_S=4$, $\tilde U=10$, and $\varepsilon_{ph}=0.1$. The curve for $g=1.5$ is offset by $0.025$. (b) LDOS at different bias voltages $V=0$ (top), $-\varepsilon_{ph}/4$, $-\varepsilon_{ph}/2$, $-3\varepsilon_{ph}/4$, and $-\varepsilon_{ph}$ (bottom). The curves are offset for clarity. The dashed lines guide the shift of the Kondo satellites, while the circles mark their mergence. (c) Illustrations of the Kondo cotunneling processes corresponding to the Kondo peaks in (b). (d) Currents vs $\tilde\varepsilon_d$ at different bias voltages ($V$ decreases from $-\varepsilon_{ph}/4$ to $-2\varepsilon_{ph}$ by a step $\varepsilon_{ph}/4$, from top to bottom). Inset shows the corresponding differential conductance.}\label{fig2}
\end{figure}

\emph{Results and discussions.---}In the numerical results presented below, we take all the renormalized parameters to be freely tunable. $\tilde\Gamma_N$ is taken as the energy unit and the temperature is always set at zero. We consider first the phonon-assisted inelastic AT in the Kondo regime. To this end, we adopt the parameters $\tilde\varepsilon_d=-2.5$, $\tilde\Gamma_S=4$, and $\tilde U=10$ such that the Kondo effect and the on-dot paring coexist. In Fig.\,\ref{fig2}(a), it is shown that remarkable differential conductance peaks, in addition to the zero-bias Kondo peak, develop whenever the bias voltage varies by half a phonon energy. These Kondo sidebands, with their typical temperature dependence given in the Supplemental Material \cite{SM}, are consistent with the scenarios previously discussed in Figs.\,\ref{fig1}(a)-(c), and are very different from those occurring in N-QD-N systems that are separated by one phonon energy \cite{Konig1996,Chen2006,Fernandez2008,Roura2013,Rakhmilevitch2014,Roura2016}. Note also that the Kondo sidebands at positive bias are much weaker than those at negative bias, which can be ascribed to that the Kondo effect is suppressed (enhanced) at positive (negative) bias since the dot energy level gets away from (closer to) the Fermi level of lead N. Furthermore, as compared with the Kondo resonance at zero EPI [see the red dashed curve in Fig.\,\ref{fig2}(a)], the zero-bias peak at finite EPI is significantly reduced and narrowed.

The underlying physics about why the conductance peaks are separated by $\varepsilon_{ph}/2$ can be acquired by examining the LDOS, since the conductance from the Kondo-Andreev tunneling processes is roughly proportional to the convolution of electron and hole density of states \cite{Deacon2010b,Chevallier2011}. Fig.\,\ref{fig2}(b) presents the LDOS for several bias voltages decreasing in a step of $\varepsilon_{ph}/4$. In equilibrium, multiple Kondo satellites ($\omega=n\varepsilon_{ph}$) exhibit on each side of the main Kondo resonance ($\omega=0$) due to the EPI. In the following, we will focus on the nearest two satellites around the main resonance. In nonequilibrium, the main resonance and the two Kondo satellites all split into two subpeaks, resulting in totally six Kondo peaks in the LDOS as indicated by $\textrm{L}_1$ ($\omega=V$), $\textrm{L}_2$ ($\omega=V-\varepsilon_{ph}$), $\textrm{L}_3$ ($\omega=V+\varepsilon_{ph}$), $\textrm{R}_1$ ($\omega=-V$), $\textrm{R}_2$ ($\omega=-V-\varepsilon_{ph}$), and $\textrm{R}_3$ ($\omega=-V+\varepsilon_{ph}$) in Fig.\,\ref{fig2}(b). When the bias is tuned to $V=-\varepsilon_{ph}/2$, the two peaks $\textrm{L}_1$ and $\textrm{R}_2$, as well as $\textrm{L}_3$ and $\textrm{R}_1$, merge into a single pronounced resonance (marked by red circles), respectively. Clearly, the convolution of these two merged Kondo resonances is larger than the convolution of $\textrm{L}_1$ and $\textrm{R}_1$ at $V=-\varepsilon_{ph}/4$, thereby cooperatively giving rise to a conductance peak at $V=-\varepsilon_{ph}/2$. Similarly, at $V=-\varepsilon_{ph}$, the two Kondo satellites $\textrm{L}_3$ and $\textrm{R}_2$ get merged at $\omega=0$ and thus results in a conductance peak. In short words, the Kondo sidebands always appear in the conductance at the bias voltage $V$ under which the LDOS exhibits Kondo-peak cooperative enhancement within the bias window $\omega\in[-V,\,V]$.

The cotunneling processes associated with some Kondo peaks in the LDOS are illustrated in Fig.\,\ref{fig2}(c). It is shown that the cotunneling processes of the $\textrm{L}_i\,(i=1,2,3)$ and $\textrm{R}_i$ Kondo peaks are of the second and fourth order, respectively. This explains why the $\textrm{L}_i$ Kondo resonances are stronger than the $\textrm{R}_i$ resonances. Specifically, in the Kondo process of $\textrm{L}_1$, a localized spin-$\uparrow$ electron tunnels out to lead N, followed closely by a spin-$\downarrow$ electron at $\mu_N$ tunneling into the QD. At low temperatures, a coherent superposition of such second-order spin-flip cotunneling events yields a many-body spin singlet comprising of the localized and N-lead electrons, which manifests itself as the sharp Kondo resonance $\textrm{L}_1$ in the LDOS. When $\tilde\Gamma_S>\tilde\Gamma_N$ as in our case, the AT can also take part in the Kondo cotunneling process. For example, in the Kondo-Andreev process of $\textrm{R}_1$, the localized spin-$\uparrow$ electron first tunnels out to $\mu_N$ and a Cooper pair in the S lead splits into two electrons with opposite spins. The split spin-$\downarrow$ electron then tunnels into the QD while the other electron transfer through the QD to the empty state with energy $-V$ in lead N. The coherent superposition of such fourth-order spin-flip cotunneling events leads to the weak Kondo resonance $\textrm{R}_1$ in the LDOS \cite{Fazio1998,Sun2001,Yamada2011}. Other Kondo peaks such as $\textrm{L}_2$, $\textrm{L}_3$, $\textrm{R}_2$, and $\textrm{R}_3$ are produced by similar Kondo and Kondo-Andreev cotunneling processes but with one phonon being emitted.

\begin{figure}[t]
\centering
\includegraphics[width=0.95\columnwidth]{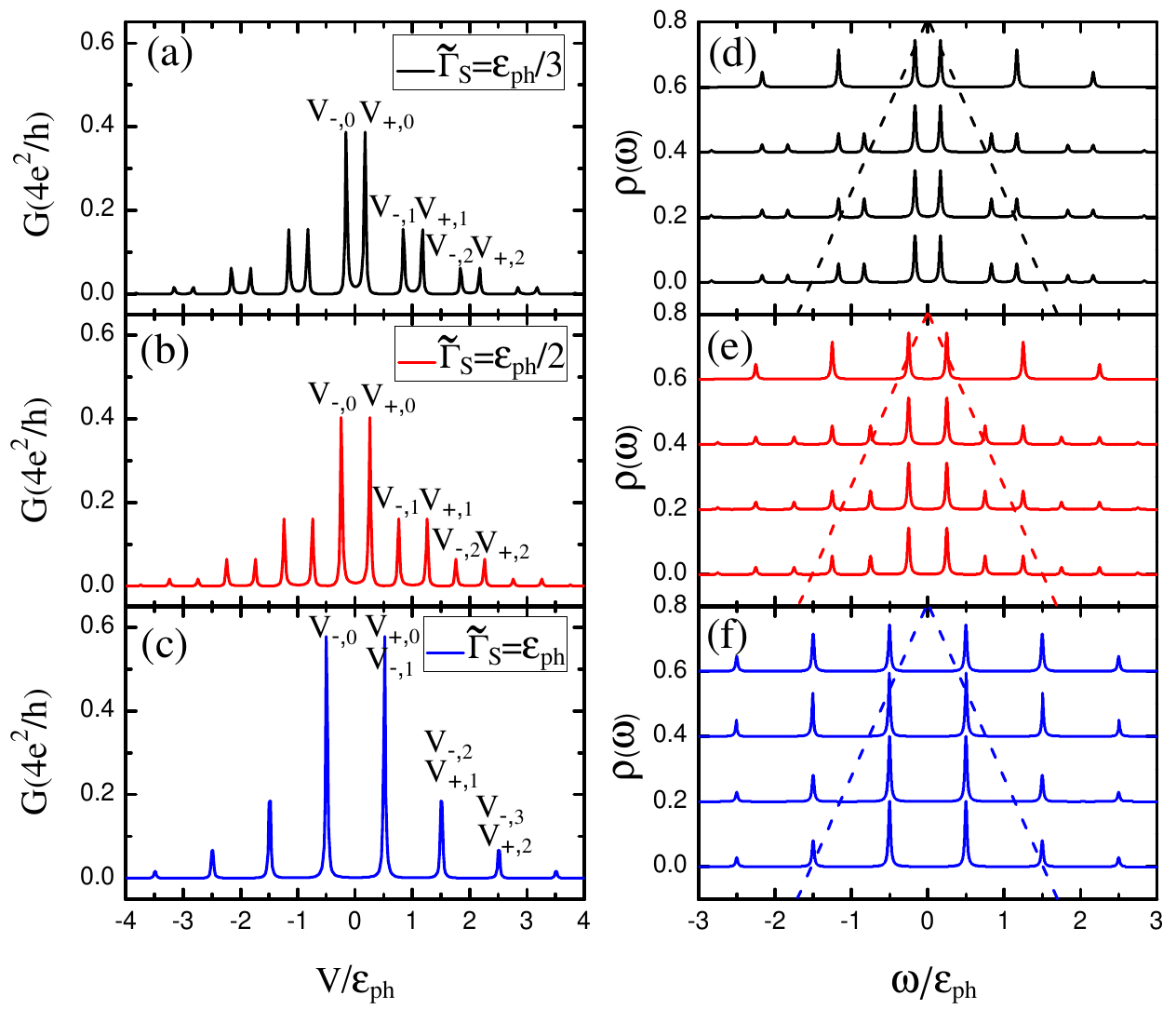}
\caption{(Color online) Left panel: the conductance of a noninteracting ($\tilde U=0$) N-QD-S system with finite EPI $g=0.8$ for different $\tilde\Gamma_S$ as indicated. Right panel: the corresponding LDOS at four different bias voltages $V=0$, $\varepsilon_{ph}/2$, $\varepsilon_{ph}$, and $3\varepsilon_{ph}/2$ (from top to bottom in each figure). The curves are offset for clarity. The horizontal distance between the two dashed lines represents the bias window $\omega\in[-V,\,V]$. Other parameters are $\tilde\varepsilon_d=0$ and $\varepsilon_{ph}=30$.}\label{fig3}
\end{figure}

The current and conductance as a function of the dot level $\tilde\varepsilon_d$ are further investigated [Fig.\,\ref{fig2}(d)]. As we can see, both quantities change monotonously with $\tilde\varepsilon_d$. This is different from those in conventional AT regime where characteristic peaks show up whenever the the dot level $\tilde\varepsilon_d$ changes by $\varepsilon_{ph}/2$ \cite{Zhang2012,Gramich2015}. The featureless nature of our $I$ vs $\tilde\varepsilon_d$ and $G$ vs $\tilde\varepsilon_d$ curves can be readily understood. As long as $\tilde\varepsilon_d$ is always restricted in the Kondo regime, the resulting Kondo resonances are robust and no additional phonon-assisted channel could be opened or closed when $\tilde\varepsilon_d$ is varied.

For comparison, we now turn to investigate the conventional inelastic AT in a noninteracting ($\tilde U=0$) N-QD-S with the QD level $\tilde\varepsilon_d=0$ fixed at the Fermi energy. In this parameter regime, the two ABSs appear at $\pm E_A$ with $E_A=\tilde\Gamma_S/2$. In Figs.\,\ref{fig3}(a)-3(c), the conductance is displayed for three values of $\tilde\Gamma_S$. Different from the conductance behaviors in the Kondo regime, there are indeed two sets of phonon sidebands at $V_{\pm,n}\equiv \pm E_A+n\varepsilon_{ph}$, each separated by one phonon energy, in agreement with our previous discussions of Figs.\,\ref{fig1}(d)-1(f). Generally, the two sets of sidebands are interleaved [Figs.\,3(a) and 3(b)]. For $E_A=\varepsilon_{ph}/2$ [Fig.\,3(c)], the two sets of sidebands merge with each others. This corresponds to the $I$-$V$ staircases addressed previously \cite{Zhang2012}. These conductance behaviors displayed can also be traced back to the LDOS at different bias voltages, as shown in Figs.\,\ref{fig3}(d)-3(f). At zero bias, only hole-type (electron-type) sidebands of the upper (lower) ABS appear at $E_A+n\varepsilon_{ph}$ ($-E_A-n\varepsilon_{ph}$), with $n>0$, which can be attributed to the fact that the upper (lower) ABS is fully empty (occupied) and the phonon absorption is unavailable at zero temperature \cite{Chen2005}. For finite bias larger than $E_A$, the upper ABS becomes occupied, therefore phonon sidebands develop on both sides of each ABS. Upon adjusting $\tilde\Gamma_S$ such that $E_A=n\varepsilon_{ph}/2$ the sidebands associated with the two ABSs merge together [see Fig.\,\ref{fig3}(f)]. For a weak but nonzero $\tilde U$, the ground state is still a BCS singlet as $\tilde U=0$. Only the height and distance between the two ABSs are slightly affected, while the general scenario of the phonon sidebands remains unchanged \cite{Baranski2015a,SM}.

\emph{Conclusions.---}We have predicted in N-QD-S systems a series of differential conductance subpeaks developed at $V=n\varepsilon_{ph}/2$ and resulting from phonon-assisted inelastic Kondo-Andreev cotunnelings. These structure are truly remarkable when compared with the transport characteristics of i) the conventional inelastic AT in N-QD-S systems \cite{Zhang2012,Baranski2015a,Gramich2015} and ii) the inelastic Kondo cotunneling in the N-QD-N systems \cite{Konig1996,Chen2006,Fernandez2008,Roura2013,Rakhmilevitch2014,Roura2016}. Our prediction might be observed in the carbon nanotube device fabricated by J. Gramich \emph{et al.}\,\cite{Gramich2015} as long as the Kondo regime is achieved at low temperatures. Similar phenomena can also be expected when the device is driven by a microwave \cite{Cho1999} instead of the electron-phonon coupling.

\section{Acknowledgments}
This work was supported by NSFC (Grant Nos.\,11325417, 11674139, and 11574007) and NBRP of China (2015CB921102).

\end{document}

\title{Supplemental material for: Inelastic Kondo-Andreev tunnelings in a vibrating quantum dot}

\maketitle
\setcounter{equation}{0}
\setcounter{figure}{0}
\makeatletter
\renewcommand{\theequation}{S\arabic{equation}}
\renewcommand{\thefigure}{S\arabic{figure}}
\onecolumngrid
In Sec.\,\ref{EOM}, we first present the detailed derivation of the equations of motion (EOMs) and the decoupling approximations we made. Then the effects of bias voltage and temperature on the local density of states (LDOS) and the differential conductance are given as a supplement to the main text. In Sec.\,\ref{SOPT}, the modified second-order perturbation theory (SOPT) in the Coulomb interaction proposed recently in Ref.\,[\onlinecite{Yamada2011}] is employed to calculate the LDOS and the differential conductance, which agrees with and complements the equation-of-motion results in the main text.

\section{Equation-of-motion method}\label{EOM}
\subsection{Detailed derivation of the equations of motion and the decoupling approximations}
Our starting point is the transformed Hamiltonian in the main text
\begin{equation}
\tilde H=\sum_{k,\sigma,\beta}\varepsilon_{k}c_{k\sigma\beta}^{\dag}c_{k\sigma\beta}-\Delta\sum_{k,\beta}\delta_{\beta,S}(  c_{k\uparrow\beta}^{\dag
}c_{-k\downarrow\beta}^{\dag}+c_{-k\downarrow\beta}c_{k\uparrow\beta})+\sum_{\sigma}\tilde{\varepsilon}_{d}d_{\sigma}^{\dag}d_{\sigma}+\tilde
{U}n_{d\uparrow}n_{d\downarrow}+\sum_{k,\sigma,\beta}\tilde{V}_{\beta}(c_{k\sigma\beta}^{\dag}d_{\sigma}+d_{\sigma}^{\dag}c_{k\sigma\beta}).\label{A1}
\end{equation}
Using the Zubarev notation \cite{Zubarev1960}, the Fourier transform of the EOM for the retarded Green's function (GF) involving fermionic operators $A$ and $B$ can be written as
\begin{equation}
(\omega+i\delta)\langle\langle A;B\rangle\rangle_\omega=\langle[A,B]_+\rangle+\langle\langle [A,\tilde H]_-;B\rangle\rangle_\omega,
\end{equation}
where $\delta\rightarrow0^+$ and the subscript $\pm$ stands for the anticommutation (commutation) relationship. In the following, the imaginary part $i\delta$ going alongside $\omega$ as well as the subscript $\omega$ in the GFs will be implicit to simplify the notations. Firstly, the EOMs for the dot GFs are
\begin{equation}
(  \omega-\tilde{\varepsilon}_{d})  \langle \langle d_{\sigma};\hat{O}\rangle \rangle=\langle [  d_{\sigma},\hat{O}]  _{+}\rangle +\sum_{k,\beta}\tilde{V}_{\beta}\langle
\langle c_{k\sigma\beta};\hat{O}\rangle \rangle +\tilde{U}\langle \langle n_{d\bar{\sigma}}d_{\sigma};\hat{O}\rangle\rangle,\label{A2}
\end{equation}
\begin{equation}
(  \omega+\tilde{\varepsilon}_{d})  \langle \langle d_{\bar{\sigma}}^{\dag};\hat{O}\rangle \rangle =\langle[  d_{\bar{\sigma}}^{\dag},\hat{O}]  _{+}\rangle
-\sum_{k,\beta}\tilde{V}_{\beta}\langle \langle c_{-k\bar{\sigma}\beta}^{\dag};\hat{O}\rangle \rangle -\tilde{U}\langle\langle n_{d\sigma}d_{\bar{\sigma}}^{\dag};\hat{O}\rangle\rangle, \label{A3}
\end{equation}
where the operator $\hat{O}$ is either $d_{\sigma}^{\dag}$ or $d_{\bar{\sigma}}$. The EOMs for the GFs involving lead electrons in Eqs.\,(\ref{A2}) and (\ref{A3}) are given by
\begin{equation}
(  \omega-\varepsilon_{k})  \langle \langle c_{k\sigma\beta};\hat{O}\rangle \rangle =-\sigma\Delta\delta_{\beta,S}\langle \langle c_{-k\bar{\sigma}\beta}^{\dag};\hat{O}\rangle \rangle +\tilde{V}_{\beta}\langle \langle d_{\sigma};\hat{O}\rangle \rangle, \label{A4}
\end{equation}
\begin{equation}
(  \omega+\varepsilon_{-k})  \langle \langle c_{-k\bar{\sigma}\beta}^{\dag};\hat{O}\rangle \rangle =-\sigma\Delta\delta_{\beta,S}\langle \langle c_{k\sigma\beta};\hat{O}\rangle
\rangle -\tilde{V}_{\beta}\langle \langle d_{\bar{\sigma}}^{\dag};\hat{O}\rangle \rangle, \label{A5}
\end{equation}
where $\sigma$ ($\bar\sigma$) in the subscript represents the spin orientation $\uparrow$ ($\downarrow$) or $\downarrow$ ($\uparrow$), while those appearing in the coefficients are set to be $\pm 1$ for $\uparrow$ ($\downarrow$).
Substituting Eqs.\,(\ref{A4}) and (\ref{A5}) into Eqs.\,(\ref{A2}) and (\ref{A3}) one can obtain
\begin{eqnarray}
&&[  \omega-\tilde{\varepsilon}_{d}-\sum_{k,\beta}\frac{\tilde{V}_{\beta}^{2}(  \omega+\varepsilon_{-k})  }{(  \omega-\varepsilon_{k})
(  \omega+\varepsilon_{-k})  -\delta_{\beta,S}\Delta^{2}}]\langle \langle d_{\sigma};\hat{O}\rangle \rangle\notag\\
&&=\langle [  d_{\sigma},\hat{O}]  _{+}\rangle+\sum_{k,\beta}\frac{\tilde{V}_{\beta}^{2}\sigma\Delta\delta_{\beta,S}}{(  \omega-\varepsilon_{k})  (  \omega+\varepsilon_{-k})-\delta_{\beta,S}\Delta^{2}}\langle \langle d_{\bar{\sigma}}^{\dag};\hat{O}\rangle \rangle +\tilde{U}\langle \langle n_{d\bar{\sigma}}d_{\sigma};\hat{O}\rangle \rangle,\label{A6}
\end{eqnarray}
\begin{eqnarray}
&&[  \omega+\tilde{\varepsilon}_{d}-\sum_{k,\beta}\frac{\tilde{V}_{\beta}^{2}(  \omega-\varepsilon_{k})  }{(  \omega-\varepsilon_{k})(  \omega+\varepsilon_{-k})  -\delta_{\beta,S}\Delta^{2}}]
\langle \langle d_{\bar{\sigma}}^{\dag};\hat{O}\rangle\rangle\notag\\
&&=\langle [  d_{\bar{\sigma}}^{\dag},\hat{O}]_{+}\rangle +\sum_{k,\beta}\frac{\tilde{V}_{\beta}^{2}\sigma\Delta\delta_{\beta,S}}{(  \omega-\varepsilon_{k})  (  \omega+\varepsilon
_{-k})  -\delta_{\beta,S}\Delta^{2}}\langle \langle d_{\sigma};\hat{O}\rangle \rangle -\tilde{U}\langle \langle n_{d\sigma}d_{\bar{\sigma}}^{\dag};\hat{O}\rangle \rangle.\label{A7}
\end{eqnarray}
We now turn to the interacting part of the EOMs concerning the GFs $\langle \langle n_{d\bar{\sigma}}d_{\sigma};\hat{O}\rangle \rangle $ and $\langle \langle n_{d\sigma}d_{\bar{\sigma}}^{\dag};\hat
{O}\rangle \rangle $ that appearing on the right-hand side of Eqs.\,(\ref{A2}) and (\ref{A3}). The EOMs are
\begin{equation}
(  \omega-\tilde{\varepsilon}_{d}-\tilde{U})  \langle \langle n_{d\bar{\sigma}}d_{\sigma};\hat{O}\rangle \rangle =\langle[  n_{d\bar{\sigma}}d_{\sigma},\hat{O}]  _{+}\rangle
-\sum_{k,\beta}\tilde{V}_{\beta}(  \langle \langle d_{\bar{\sigma}}^{\dag}d_{\sigma}c_{-k\bar{\sigma}\beta};\hat{O}\rangle\rangle -\langle \langle n_{d\bar{\sigma}}c_{k\sigma\beta};\hat{O}\rangle \rangle +\langle \langle c_{-k\bar{\sigma}\beta}^{\dag}d_{\bar{\sigma}}d_{\sigma};\hat{O}\rangle\rangle ),\label{A8}
\end{equation}
\begin{equation}
(  \omega+\tilde{\varepsilon}_{d}+\tilde{U})  \langle \langle n_{d\sigma}d_{\bar{\sigma}}^{\dag};\hat{O}\rangle \rangle
=\langle [  n_{d\sigma}d_{\bar{\sigma}}^{\dag},\hat{O}]_{+}\rangle +\sum_{k,\beta}\tilde{V}_{\beta}(  \langle\langle c_{k\sigma\beta}^{\dag}d_{\bar{\sigma}}^{\dag}d_{\sigma};\hat
{O}\rangle \rangle -\langle \langle c_{-k\bar{\sigma}\beta}^{\dag}n_{d\sigma};\hat{O}\rangle \rangle -\langle\langle d_{\sigma}^{\dag}d_{\bar{\sigma}}^{\dag}c_{k\sigma\beta};\hat{O}\rangle \rangle ),\label{A9}
\end{equation}
and each of them generates three new GFs whose EOMs are
\begin{eqnarray}
(\omega-\varepsilon_{-k})  \langle \langle d_{\bar{\sigma}}^{\dag}d_{\sigma}c_{-k\bar{\sigma}\beta};\hat{O}\rangle &&=\langle [  d_{\bar{\sigma}}^{\dag}d_{\sigma}c_{-k\bar{\sigma}\beta
},\hat{O}]  _{+}\rangle +\sigma\Delta\delta_{\beta,S}\langle\langle c_{k\sigma\beta}^{\dag}d_{\bar{\sigma}}^{\dag}d_{\sigma};\hat{O}\rangle \rangle -\tilde{V}_{\beta}\langle \langle n_{d\bar{\sigma}}d_{\sigma};\hat{O}\rangle \rangle \notag\\
&&-\sum_{k^{\prime},\beta^{\prime}}\tilde{V}_{\beta^{\prime}}\langle\langle d_{\bar{\sigma}}^{\dag}c_{-k\bar{\sigma}\beta}c_{k^{\prime}\sigma\beta^{\prime}};\hat{O}\rangle \rangle +\sum_{k^{\prime
},\beta^{\prime}}\tilde{V}_{\beta^{\prime}}\langle \langle c_{-k^{\prime}\bar{\sigma}\beta^{\prime}}^{\dag}c_{-k\bar{\sigma}\beta}d_{\sigma};\hat{O}\rangle \rangle,\label{A10}
\end{eqnarray}
\begin{eqnarray}
(  \omega-\varepsilon_{k})  \langle \langle n_{d\bar{\sigma}}c_{k\sigma\beta};\hat{O}\rangle \rangle &&=\langle [n_{d\bar{\sigma}}c_{k\sigma\beta},\hat{O}]  _{+}\rangle
-\sigma\Delta\delta_{\beta,S}\langle \langle c_{-k\bar{\sigma}\beta}^{\dag}n_{d\bar{\sigma}};\hat{O}\rangle \rangle +\tilde{V}_{\beta}\langle \langle n_{d\bar{\sigma}}d_{\sigma};\hat{O}\rangle \rangle \notag\\
&&+\sum_{k^{\prime},\beta^{\prime}}\tilde{V}_{\beta^{\prime}}\langle \langle d_{\bar{\sigma}}^{\dag}c_{-k^{\prime}\bar{\sigma}\beta^{\prime}}c_{k\sigma\beta};\hat{O}\rangle \rangle +\sum_{k^{\prime},\beta^{\prime}}\tilde{V}
_{\beta^{\prime}}\langle \langle c_{-k^{\prime}\bar{\sigma}\beta^{\prime}}^{\dag}c_{k\sigma\beta}d_{\bar{\sigma}};\hat{O}\rangle\rangle,\label{A12}
\end{eqnarray}
\begin{eqnarray}
(  \omega+\varepsilon_{-k}^{-})  \langle \langle
c_{-k\bar{\sigma}\beta}^{\dag}d_{\bar{\sigma}}d_{\sigma};\hat{O}\rangle\rangle &&=\langle [  c_{-k\bar{\sigma}\beta}^{\dag}d_{\bar{\sigma}}d_{\sigma},\hat{O}]  _{+}\rangle -\sigma
\Delta\delta_{\beta,S}\langle \langle c_{k\sigma\beta}d_{\bar{\sigma}}d_{\sigma};\hat{O}\rangle \rangle -\tilde{V}_{\beta}\langle \langle n_{d\bar{\sigma}}d_{\sigma};\hat{O}\rangle\rangle\notag\\
&&-\sum_{k^{\prime},\beta^{\prime}}\tilde{V}_{\beta^{\prime}}\langle \langle d_{\bar{\sigma}}c_{-k\bar{\sigma}\beta}^{\dag}c_{k^{\prime}\sigma\beta^{\prime}};\hat{O}\rangle \rangle
+\sum_{k^{\prime},\beta^{\prime}}\tilde{V}_{\beta^{\prime}}\langle\langle c_{-k\bar{\sigma}\beta}^{\dag}c_{-k^{\prime}\bar{\sigma}\beta^{\prime}}d_{\sigma};\hat{O}\rangle \rangle,\label{A14}
\end{eqnarray}
and
\begin{eqnarray}
(\omega+\varepsilon_{k})  \langle \langle c_{k\sigma\beta}^{\dag}d_{\bar{\sigma}}^{\dag}d_{\sigma};\hat{O}\rangle \rangle
&&=\langle [  c_{k\sigma\beta}^{\dag}d_{\bar{\sigma}}^{\dag}d_{\sigma},\hat{O}]  _{+}\rangle +\sigma\Delta\delta_{\beta,S}\langle\langle d_{\bar{\sigma}}^{\dag}d_{\sigma}c_{-k\bar{\sigma}\beta};\hat
{O}\rangle \rangle +\tilde{V}_{\beta}\langle \langle n_{d\sigma}d_{\bar{\sigma}}^{\dag};\hat{O}\rangle \rangle\notag\\
&&-\sum_{k^{\prime},\beta^{\prime}}\tilde{V}_{\beta^{\prime}}\langle\langle d_{\bar{\sigma}}^{\dag}c_{k\sigma\beta}^{\dag}c_{k^{\prime}\sigma\beta^{\prime}};\hat{O}\rangle \rangle -\sum_{k^{\prime
},\beta^{\prime}}\tilde{V}_{\beta^{\prime}}\langle \langle c_{k\sigma\beta}^{\dag}c_{-k^{\prime}\bar{\sigma}\beta^{\prime}}^{\dag}d_{\sigma};\hat{O}\rangle \rangle,\label{A11}
\end{eqnarray}
\begin{eqnarray}
(  \omega+\varepsilon_{-k})  \langle \langle c_{-k\bar{\sigma}\beta}^{\dag}n_{d\sigma};\hat{O}\rangle \rangle &&=\langle[  c_{-k\bar{\sigma}\beta}^{\dag}n_{d\sigma},\hat{O}]
_{+}\rangle -\sigma\Delta\delta_{\beta,S}\langle \langle n_{d\sigma}c_{k\sigma\beta};\hat{O}\rangle \rangle -\tilde{V}_{\beta}\langle \langle n_{d\sigma}d_{\bar{\sigma}}^{\dag};\hat{O}\rangle \rangle\notag\\
&& -\sum_{k^{\prime},\beta^{\prime}}\tilde{V}_{\beta^{\prime}}\langle \langle d_{\sigma}^{\dag}c_{-k\bar{\sigma}\beta}^{\dag}c_{k^{\prime}\sigma\beta^{\prime}};\hat{O}\rangle
\rangle+\sum_{k^{\prime},\beta^{\prime}}\tilde{V}_{\beta^{\prime}}\langle \langle c_{k^{\prime}\sigma\beta^{\prime}}^{\dag}c_{-k\bar{\sigma}\beta}^{\dag}d_{\sigma};\hat{O}\rangle \rangle,\label{A16}
\end{eqnarray}
\begin{eqnarray}
(  \omega-\varepsilon_{k}^{-})  \langle \langle d_{\sigma}^{\dag}d_{\bar{\sigma}}^{\dag}c_{k\sigma\beta};\hat{O}\rangle\rangle &&=\langle [  d_{\sigma}^{\dag}d_{\bar{\sigma}}^{\dag
}c_{k\sigma\beta},\hat{O}]  _{+}\rangle -\sigma\Delta\delta_{\beta,S}\langle \langle d_{\sigma}^{\dag}d_{\bar{\sigma}}^{\dag}c_{-k\bar{\sigma}\beta}^{\dag};\hat{O}\rangle \rangle -\tilde
{V}_{\beta}\langle \langle n_{d\sigma}d_{\bar{\sigma}}^{\dag};\hat{O}\rangle \rangle\notag\\
&& -\sum_{k^{\prime},\beta^{\prime}}\tilde{V}_{\beta^{\prime}}\langle \langle d_{\sigma}^{\dag}c_{-k^{\prime}\bar{\sigma}\beta^{\prime}}^{\dag}c_{k\sigma\beta};\hat{O}\rangle
\rangle +\sum_{k^{\prime},\beta^{\prime}}\tilde{V}_{\beta^{\prime}}\langle \langle c_{k^{\prime}\sigma\beta^{\prime}}^{\dag}c_{k\sigma\beta}d_{\bar{\sigma}}^{\dag};\hat{O}\rangle \rangle,\label{A18}
\end{eqnarray}
where $\varepsilon_{\pm k}^{-}=\varepsilon_{\pm k}-(  2\tilde{\varepsilon}_{d}+\tilde{U})  $. Now four more new GFs which involve one lead operator and three dot operators are generated. The EOMs of them are
\begin{eqnarray}
(  \omega+\varepsilon_{-k})  \langle \langle c_{-k\bar{\sigma}\beta}^{\dag}n_{d\bar{\sigma}};\hat{O}\rangle \rangle
&&=\langle [  c_{-k\bar{\sigma}\beta}^{\dag}n_{d\bar{\sigma}},\hat{O}]  _{+}\rangle -\sigma\Delta\delta_{\beta,S}\langle\langle n_{d\bar{\sigma}}c_{k\sigma\beta};\hat{O}\rangle\rangle\notag\\
&& -\sum_{k^{\prime},\beta^{\prime}}\tilde{V}_{\beta^{\prime}}\langle \langle d_{\bar{\sigma}}^{\dag}c_{-k\bar{\sigma}\beta}^{\dag}c_{-k^{\prime}\bar{\sigma}\beta^{\prime}};\hat{O}\rangle\rangle -\sum_{k^{\prime},\beta^{\prime}}\tilde{V}_{\beta^{\prime}}\langle \langle c_{-k\bar{\sigma}\beta}^{\dag}c_{-k^{\prime}\bar{\sigma}\beta^{\prime}}^{\dag}d_{\bar{\sigma}};\hat{O}\rangle\rangle,\label{A13}
\end{eqnarray}
\begin{eqnarray}
(  \omega-\varepsilon_{k}^{+})  \langle \langle c_{k\sigma\beta}d_{\bar{\sigma}}d_{\sigma};\hat{O}\rangle \rangle
&&=\langle [  c_{k\sigma\beta}d_{\bar{\sigma}}d_{\sigma},\hat{O}]  _{+}\rangle -\sigma\Delta\delta_{\beta,S}\langle\langle c_{-k\bar{\sigma}\beta}^{\dag}d_{\bar{\sigma}}d_{\sigma};\hat{O}\rangle \rangle\notag\\
&&-\sum_{k^{\prime},\beta^{\prime}}\tilde{V}_{\beta^{\prime}}\langle \langle d_{\bar{\sigma}}c_{k\sigma\beta}c_{k^{\prime}\sigma\beta^{\prime}};\hat{O}\rangle \rangle
-\sum_{k^{\prime},\beta^{\prime}}\tilde{V}_{\beta^{\prime}}\langle\langle c_{-k^{\prime}\bar{\sigma}\beta^{\prime}}c_{k\sigma\beta}d_{\sigma};\hat{O}\rangle \rangle,\label{A15}
\end{eqnarray}
\begin{eqnarray}
(  \omega-\varepsilon_{k})  \langle \langle n_{d\sigma}c_{k\sigma\beta};\hat{O}\rangle \rangle &&=\langle [n_{d\sigma}c_{k\sigma\beta},\hat{O}]  _{+}\rangle -\sigma
\Delta\delta_{\beta,S}\langle \langle c_{-k\bar{\sigma}\beta}^{\dag}n_{d\sigma};\hat{O}\rangle \rangle\notag\\
&& +\sum_{k^{\prime},\beta^{\prime}}\tilde{V}_{\beta^{\prime}}\langle \langle d_{\sigma}^{\dag}c_{k^{\prime}\sigma\beta^{\prime}}c_{k\sigma\beta};\hat{O}\rangle \rangle+\sum_{k^{\prime},\beta^{\prime}}\tilde{V}
_{\beta^{\prime}}\langle \langle c_{k^{\prime}\sigma\beta^{\prime}}^{\dag}c_{k\sigma\beta}d_{\sigma};\hat{O}\rangle \rangle,\label{A17}
\end{eqnarray}
\begin{eqnarray}
(  \omega+\varepsilon_{-k}^{+})  \langle \langle d_{\sigma}^{\dag}d_{\bar{\sigma}}^{\dag}c_{-k\bar{\sigma}\beta}^{\dag};\hat{O}\rangle \rangle &&=\langle [  d_{\sigma}^{\dag}
d_{\bar{\sigma}}^{\dag}c_{-k\bar{\sigma}\beta}^{\dag},\hat{O}]_{+}\rangle -\sigma\Delta\delta_{\beta,S}\langle \langle d_{\sigma}^{\dag}d_{\bar{\sigma}}^{\dag}c_{k\sigma\beta};\hat{O}\rangle\rangle\notag\\
&& -\sum_{k^{\prime},\beta^{\prime}}\tilde{V}_{\beta^{\prime}}\langle \langle d_{\sigma}^{\dag}c_{-k^{\prime}\bar{\sigma}\beta^{\prime}}^{\dag}c_{-k\bar{\sigma}\beta}^{\dag};\hat{O}\rangle
\rangle +\sum_{k^{\prime},\beta^{\prime}}\tilde{V}_{\beta^{\prime}}\langle \langle c_{k^{\prime}\sigma\beta^{\prime}}^{\dag}c_{-k\bar{\sigma}\beta}^{\dag}d_{\bar{\sigma}}^{\dag};\hat{O}\rangle\rangle,\label{A19}
\end{eqnarray}
where $\varepsilon_{\pm k}^{+}=\varepsilon_{\pm k}+(  2\tilde{\varepsilon}_{d}+\tilde{U})  $. It is seen that many new GFs involving two lead operators and two dot operators are generated on the right-hand side of Eqs.\,(\ref{A10})-(\ref{A19}). To truncate the EOMs we make such decoupling approximations adopted by Sun \emph{et al}.\,\cite{Sun2001} as (i) $\langle \langle c_{k\sigma\beta}^{\dag}c_{k^{\prime}\sigma\beta^{\prime}}\hat{A};\hat{O}\rangle \rangle\approx\delta_{k,k^{\prime}}\delta_{\beta,\beta^{\prime}}\langle
c_{k\sigma\beta}^{\dag}c_{k\sigma\beta}\rangle \langle \langle\hat{A};\hat{O}\rangle \rangle $, $\langle \langle c_{-k^{\prime}\bar{\sigma}\beta^{\prime}}c_{k\sigma\beta}\hat{A};\hat{O}\rangle \rangle \approx\delta_{k,k^{\prime}}\delta_{\beta,\beta^{\prime}}\langle c_{-k\bar{\sigma}\beta}c_{k\sigma\beta}\rangle \langle \langle \hat{A};\hat{O}\rangle\rangle $, and $\langle \langle c_{k\sigma\beta}^{\dag
}c_{-k^{\prime}\bar{\sigma}\beta^{\prime}}^{\dag}\hat{A};\hat{O}\rangle\rangle \approx\delta_{k,k^{\prime}}\delta_{\beta,\beta^{\prime}}\langle c_{k\sigma\beta}^{\dag}c_{-k\bar{\sigma}\beta}^{\dag}\rangle\langle \langle \hat{A};\hat{O}\rangle \rangle $ with the averages $\langle c_{k\sigma\beta}^{\dag}c_{k\sigma\beta}\rangle=\langle c_{-k\bar{\sigma}\beta}^{\dag}c_{-k\bar{\sigma}\beta
}\rangle =\delta_{\beta,N}f_{N}(  \varepsilon_{k})  +\frac{1}{2}\delta_{\beta,S}\{  1-\frac{\varepsilon_{k}}{\sqrt{\varepsilon_{k}^{2}+\Delta^{2}}}[  1-2f_{S}(  \sqrt{\varepsilon_{k}^{2}
+\Delta^{2}})  ]  \}  $ and $\langle c_{-k\bar{\sigma}\beta}c_{k\sigma\beta}\rangle =\langle c_{k\sigma\beta}^{\dag}c_{-k\bar{\sigma}\beta}^{\dag}\rangle =\delta_{\beta,S}\frac
{\sigma\Delta}{2\sqrt{\varepsilon_{k}^{2}+\Delta^{2}}}[  1-2f_{S}(\sqrt{\varepsilon_{k}^{2}+\Delta^{2}})  ]  $ being obtained by the mean-filed BCS Hamiltonian. Here $\hat{A}$ is a dot operator. The other GFs such as $\langle \langle c_{k\sigma\beta}^{\dag}c_{k^{\prime}\bar{\sigma}\beta^{\prime}}\hat{A};\hat{O}\rangle \rangle \approx0$, $\langle \langle c_{k^{\prime}\sigma\beta^{\prime}}c_{k\sigma\beta}\hat{A};\hat{O}\rangle\rangle \approx0$, and $\langle \langle c_{k\sigma\beta}^{\dag}c_{k^{\prime}\sigma\beta^{\prime}}^{\dag}\hat{A};\hat{O}\rangle\rangle \approx0$. (ii) The correlation functions concerning the lead electron and dot electron are taken to be zeros. This truncation scheme is an extension of the one proposed by Meir \emph{et al}.\,\cite{Meir1991} in treating the Anderson model. The latter was used extensively to qualitatively describe the Kondo peak in the LDOS and the differential conductance (see, e.g., Refs.\,[\onlinecite{Meir1993,Swirkowicz2003,Galperin2007,Zimbovskaya2008,Monreal2009,Lim2013,Xin2015,Hoffman2015}]). Better approximations for the retarded GFs such as that derived in the Refs.\,[\onlinecite{Lacroix1981},\onlinecite{Kashcheyevs2006}] may give better results for the shape and the height of the Kondo resonance. Nevertheless, the truncation approximation we adopted in this work could at least provide the right positions of the Kondo resonances, which suffices the aim of the present work. With these approximations, one solves Eqs.\,(\ref{A10})-(\ref{A19}) to obtain the six GFs emerging on the right-hand side of Eqs.\,(\ref{A8}) and (\ref{A9}) as
\begin{eqnarray}
\langle \langle d_{\bar{\sigma}}^{\dag}d_{\sigma}c_{-k\bar{\sigma}\beta};\hat{O}\rangle \rangle
&&=-\frac{\tilde{V}_{\beta}(  \omega+\varepsilon_{k})  }{(\omega-\varepsilon_{-k})  (  \omega+\varepsilon_{k})  -\delta_{\beta,S}\Delta^{2}}\langle \langle n_{d\bar{\sigma}}d_{\sigma};\hat
{O}\rangle \rangle +\frac{\sigma\Delta\delta_{\beta,S}\tilde{V}_{\beta}}{(  \omega-\varepsilon_{-k})  (  \omega+\varepsilon_{k})  -\delta_{\beta,S}\Delta^{2}}\langle \langle n_{d\sigma}d_{\bar{\sigma}}^{\dag};\hat{O}\rangle \rangle\notag\\
&&+\frac{\tilde{V}_{\beta}\langle c_{-k\bar{\sigma}\beta}^{\dag}c_{-k\bar{\sigma}\beta}\rangle (  \omega+\varepsilon_{k})-\sigma\Delta\delta_{\beta,S}\tilde{V}_{\beta}\langle c_{k\sigma\beta
}^{\dag}c_{-k\bar{\sigma}\beta}^{\dag}\rangle }{(  \omega-\varepsilon_{-k})  (  \omega+\varepsilon_{k})  -\delta_{\beta,S}\Delta^{2}}\langle \langle d_{\sigma};\hat{O}\rangle \rangle\notag\\
&&-\frac{\tilde{V}_{\beta}\langle c_{-k\bar{\sigma}\beta}c_{k\sigma\beta}\rangle (  \omega+\varepsilon_{k})+\sigma\Delta\delta_{\beta,S}\tilde{V}_{\beta}\langle c_{k\sigma\beta}^{\dag}c_{k\sigma\beta
}\rangle }{(  \omega-\varepsilon_{-k})  (  \omega+\varepsilon_{k})  -\delta_{\beta,S}\Delta^{2}}\langle \langle d_{\bar{\sigma}}^{\dag};\hat{O}\rangle \rangle,\label{A20}
\end{eqnarray}
\begin{eqnarray}
&&\langle \langle n_{d\bar{\sigma}}c_{k\sigma\beta};\hat{O}\rangle \rangle \notag\\
&&=\frac{\tilde{V}_{\beta}\langle c_{-k\bar{\sigma}\beta}c_{k\sigma\beta}\rangle (  \omega+\varepsilon_{-k})  +\sigma\Delta\delta_{\beta,S}\tilde{V}_{\beta}\langle
c_{-k\bar{\sigma}\beta}^{\dag}c_{-k\bar{\sigma}\beta}\rangle }{(\omega-\varepsilon_{k})  (  \omega+\varepsilon_{-k})  -\delta_{\beta,S}\Delta^{2}}\langle \langle d_{\bar{\sigma}}^{\dag};\hat
{O}\rangle \rangle +\frac{\tilde{V}_{\beta}(  \omega+\varepsilon_{-k})  }{(  \omega-\varepsilon_{k})  (  \omega+\varepsilon_{-k})  -\delta_{\beta,S}\Delta^{2}}\langle \langle n_{d\bar{\sigma}}d_{\sigma};\hat{O}\rangle \rangle,\label{A22}
\end{eqnarray}
\begin{eqnarray}
&&\langle \langle c_{-k\bar{\sigma}\beta}^{\dag}d_{\bar{\sigma}}d_{\sigma};\hat{O}\rangle \rangle\notag\\
&&=\frac{\tilde{V}_{\beta}\langle c_{-k\bar{\sigma}\beta}^{\dag}c_{-k\bar{\sigma}\beta}\rangle (  \omega-\varepsilon_{k}^{+})  +\sigma\Delta\delta_{\beta,S}\tilde{V}_{\beta}\langle c_{-k\bar{\sigma}\beta
}c_{k\sigma\beta}\rangle }{(  \omega-\varepsilon_{k}^{+})  (\omega+\varepsilon_{-k}^{-})  -\delta_{\beta,S}\Delta^{2}}\langle\langle d_{\sigma};\hat{O}\rangle \rangle -\frac{\tilde
{V}_{\beta}(  \omega-\varepsilon_{k}^{+})  }{(  \omega-\varepsilon_{k}^{+})  (  \omega+\varepsilon_{-k}^{-})  -\delta_{\beta,S}\Delta^{2}}\langle \langle n_{d\bar{\sigma}}d_{\sigma};\hat{O}\rangle \rangle,\label{A24}
\end{eqnarray}
and
\begin{eqnarray}
\langle \langle c_{k\sigma\beta}^{\dag}d_{\bar{\sigma}}^{\dag}d_{\sigma};\hat{O}\rangle \rangle
&&=-\frac{\sigma\Delta\delta_{\beta,S}\tilde{V}_{\beta}}{(\omega-\varepsilon_{-k})  (  \omega+\varepsilon_{k})  -\delta_{\beta,S}\Delta^{2}}\langle \langle n_{d\bar{\sigma}}d_{\sigma};\hat
{O}\rangle \rangle +\frac{\tilde{V}_{\beta}(  \omega-\varepsilon_{-k})  }{(  \omega-\varepsilon_{-k})  (  \omega+\varepsilon_{k})  -\delta_{\beta,S}\Delta^{2}}\langle \langle n_{d\sigma}d_{\bar{\sigma}}^{\dag};\hat{O}\rangle \rangle\notag\\
&&+\frac{-\tilde{V}_{\beta}\langle c_{k\sigma\beta}^{\dag}c_{-k\bar{\sigma}\beta}^{\dag}\rangle (  \omega-\varepsilon_{-k})+\sigma\Delta\delta_{\beta,S}\tilde{V}_{\beta}\langle c_{-k\bar{\sigma
}\beta}^{\dag}c_{-k\bar{\sigma}\beta}\rangle }{(  \omega-\varepsilon_{-k})  (  \omega+\varepsilon_{k})  -\delta_{\beta,S}\Delta^{2}}\langle \langle d_{\sigma};\hat{O}\rangle \rangle\notag\\
&&-\frac{\tilde{V}_{\beta}\langle c_{k\sigma\beta}^{\dag}c_{k\sigma\beta}\rangle (  \omega-\varepsilon_{-k})  +\sigma\Delta\delta_{\beta,S}\tilde{V}_{\beta}\langle c_{-k\bar{\sigma}\beta}c_{k\sigma
\beta}\rangle }{(  \omega-\varepsilon_{-k})  (\omega+\varepsilon_{k})  -\delta_{\beta,S}\Delta^{2}}\langle\langle d_{\bar{\sigma}}^{\dag};\hat{O}\rangle \rangle,\label{A21}
\end{eqnarray}
\begin{eqnarray}
&&\langle \langle c_{-k\bar{\sigma}\beta}^{\dag}n_{d\sigma};\hat{O}\rangle \rangle\notag\\
&&=\frac{\tilde{V}_{\beta}\langle c_{k\sigma\beta}^{\dag}c_{-k\bar{\sigma}\beta}^{\dag}\rangle (\omega-\varepsilon_{k})  -\sigma\Delta\delta_{\beta,S}\tilde{V}_{\beta
}\langle c_{k\sigma\beta}^{\dag}c_{k\sigma\beta}\rangle }{(\omega-\varepsilon_{k})  (  \omega+\varepsilon_{-k})  -\delta_{\beta,S}\Delta^{2}}\langle \langle d_{\sigma};\hat{O}\rangle
\rangle -\frac{\tilde{V}_{\beta}(  \omega-\varepsilon_{k})}{(  \omega-\varepsilon_{k})  (  \omega+\varepsilon_{-k})-\delta_{\beta,S}\Delta^{2}}\langle \langle n_{d\sigma}d_{\bar{\sigma}}^{\dag};\hat{O}\rangle \rangle,\label{A23}
\end{eqnarray}
\begin{eqnarray}
&&\langle \langle d_{\sigma}^{\dag}d_{\bar{\sigma}}^{\dag}c_{k\sigma\beta};\hat{O}\rangle \rangle\notag\\
&&=\frac{\tilde{V}_{\beta}\langle c_{k\sigma\beta}^{\dag}c_{k\sigma\beta}\rangle (\omega+\varepsilon_{-k}^{+})  -\sigma\Delta\delta_{\beta,S}\tilde{V}_{\beta}\langle c_{k\sigma\beta}^{\dag}c_{-k\bar{\sigma}\beta}^{\dag
}\rangle }{(  \omega-\varepsilon_{k}^{-})  (  \omega+\varepsilon_{-k}^{+})  -\delta_{\beta,S}\Delta^{2}}\langle \langle d_{\bar{\sigma}}^{\dag};\hat{O}\rangle \rangle -\frac{\tilde{V}_{\beta}(  \omega+\varepsilon_{-k}^{+})  }{(  \omega-\varepsilon_{k}^{-})  (  \omega+\varepsilon_{-k}^{+})  -\delta_{\beta,S}\Delta^{2}}\langle \langle n_{d\sigma}d_{\bar{\sigma}}^{\dag};\hat{O}\rangle \rangle.\label{A25}
\end{eqnarray}
Substituting Eqs.\,(\ref{A20})-(\ref{A25}) into Eqs.\,(\ref{A8}) and (\ref{A9}) yields
\begin{eqnarray}
&&[  \omega-\varepsilon_{d}-\tilde{U}-S(  \omega)  ]  \langle\langle n_{d\bar{\sigma}}d_{\sigma};\hat{O}\rangle \rangle\notag\\
&&=\langle [  n_{d\bar{\sigma}}d_{\sigma},\hat{O}]_{+}\rangle -M(  \omega)  \langle \langle d_{\sigma};\hat{O}\rangle \rangle +K(  \omega)  \langle\langle d_{\bar{\sigma}}^{\dag};\hat{O}\rangle \rangle-\sum_{k,\beta}\frac{\tilde{V}_{\beta}^{2}\sigma\Delta\delta_{\beta,S}}{(  \omega-\varepsilon_{-k})  (  \omega+\varepsilon_{k})
-\delta_{\beta,S}\Delta^{2}}\langle \langle n_{d\sigma}d_{\bar{\sigma}}^{\dag};\hat{O}\rangle \rangle,\label{A26}
\end{eqnarray}
\begin{eqnarray}
&&[  \omega+\varepsilon_{d}+\tilde{U}-T(  \omega)  ]  \langle\langle n_{d\sigma}d_{\bar{\sigma}}^{\dag};\hat{O}\rangle\rangle\notag\\
&&=\langle [  n_{d\sigma}d_{\bar{\sigma}}^{\dag},\hat{O}]  _{+}\rangle -N(  \omega)  \langle \langle d_{\bar{\sigma}}^{\dag};\hat{O}\rangle \rangle +L(  \omega)\langle \langle d_{\sigma};\hat{O}\rangle \rangle-\sum_{k,\beta}\frac{\tilde{V}_{\beta}^{2}\sigma\Delta\delta_{\beta,S}}{(  \omega-\varepsilon_{-k})  (  \omega+\varepsilon_{k})-\delta_{\beta,S}\Delta^{2}}\langle \langle n_{d\bar{\sigma}}d_{\sigma};\hat{O}\rangle \rangle,\label{A27}
\end{eqnarray}
where
\begin{eqnarray}
S(  \omega) &&=\sum_{k,\beta}\frac{\tilde{V}_{\beta}^{2}(\omega+\varepsilon_{k})  }{(  \omega-\varepsilon_{-k})  (\omega+\varepsilon_{k})  -\delta_{\beta,S}\Delta^{2}}+\sum_{k,\beta}
\frac{\tilde{V}_{\beta}^{2}(  \omega+\varepsilon_{-k})  }{(\omega-\varepsilon_{k})  (  \omega+\varepsilon_{-k})  -\delta_{\beta,S}\Delta^{2}}+\sum_{k,\beta}\frac{\tilde{V}_{\beta}^{2}(  \omega-\varepsilon_{k}^{+})  }{(  \omega-\varepsilon_{k}^{+})  (\omega+\varepsilon_{-k}^{-})  -\delta_{\beta,S}\Delta^{2}},\notag\\
\label{A28}
\end{eqnarray}
\begin{eqnarray}
M(  \omega) && =\sum_{k,\beta}\tilde{V}_{\beta}^{2}\frac{\langle c_{-k\bar{\sigma}\beta}^{\dag}c_{-k\bar{\sigma}\beta}\rangle (\omega+\varepsilon_{k})  -\sigma\Delta\delta_{\beta,S}\langle
c_{k\sigma\beta}^{\dag}c_{-k\bar{\sigma}\beta}^{\dag}\rangle }{(\omega-\varepsilon_{-k})  (  \omega+\varepsilon_{k})  -\delta_{\beta,S}\Delta^{2}}+\sum_{k,\beta}\tilde{V}_{\beta}^{2}\frac{\langle
c_{-k\bar{\sigma}\beta}^{\dag}c_{-k\bar{\sigma}\beta}\rangle (\omega-\varepsilon_{k}^{+})  +\sigma\Delta\delta_{\beta,S}\langle c_{-k\bar{\sigma}\beta}c_{k\sigma\beta}\rangle }{(  \omega-\varepsilon
_{k}^{+})  (  \omega+\varepsilon_{-k}^{-})  -\delta_{\beta,S}\Delta^{2}},\notag\\
\label{A29}
\end{eqnarray}
\begin{eqnarray}
K(  \omega)&&=\sum_{k,\beta}\tilde{V}_{\beta}^{2}\frac{\langle c_{-k\bar{\sigma}\beta}c_{k\sigma\beta}\rangle (  \omega+\varepsilon_{k})  +\sigma\Delta\delta_{\beta,S}\langle c_{k\sigma\beta}^{\dag}c_{k\sigma\beta}\rangle }{(  \omega-\varepsilon_{-k})  (\omega+\varepsilon_{k})  -\delta_{\beta,S}\Delta^{2}}+\sum_{k,\beta}\tilde{V}_{\beta}^{2}\frac{\langle c_{-k\bar{\sigma}\beta}c_{k\sigma\beta}\rangle (  \omega+\varepsilon_{-k})  +\sigma\Delta\delta_{\beta,S}\langle c_{-k\bar{\sigma}\beta}^{\dag}c_{-k\bar{\sigma}\beta}\rangle }{(  \omega-\varepsilon_{k})  (  \omega+\varepsilon_{-k})  -\delta_{\beta,S}\Delta^{2}},\notag\\
\label{A30}
\end{eqnarray}
and
\begin{eqnarray}
T(  \omega)  &&=\sum_{k,\beta}\frac{\tilde{V}_{\beta}^{2}(\omega-\varepsilon_{-k})  }{(  \omega-\varepsilon_{-k})  (\omega+\varepsilon_{k})  -\delta_{\beta,S}\Delta^{2}}+\sum_{k,\beta}\frac{\tilde{V}_{\beta}^{2}(  \omega-\varepsilon_{k})  }{(\omega-\varepsilon_{k})  (  \omega+\varepsilon_{-k})  -\delta_{\beta,S}\Delta^{2}}+\sum_{k,\beta}\frac{\tilde{V}_{\beta}^{2}(  \omega+\varepsilon
_{-k}^{+})  }{(  \omega-\varepsilon_{k}^{-})  (\omega+\varepsilon_{-k}^{+})  -\delta_{\beta,S}\Delta^{2}},\notag\\
\label{A31}
\end{eqnarray}
\begin{eqnarray}
N(  \omega)  &&=\sum_{k,\beta}\tilde{V}_{\beta}^{2}\frac{\langle c_{k\sigma\beta}^{\dag}c_{k\sigma\beta}\rangle (  \omega+\varepsilon_{-k}^{+})  -\sigma\Delta\delta_{\beta,S}\langle c_{k\sigma\beta}^{\dag}c_{-k\bar{\sigma}\beta}^{\dag}\rangle }{(  \omega-\varepsilon_{k}^{-})  (  \omega+\varepsilon_{-k}^{+})  -\delta_{\beta,S}\Delta^{2}}+\sum_{k,\beta}\tilde{V}_{\beta}^{2}\frac{\langle
c_{k\sigma\beta}^{\dag}c_{k\sigma\beta}\rangle (  \omega-\varepsilon_{-k})  +\sigma\Delta\delta_{\beta,S}\langle c_{-k\bar{\sigma}
\beta}c_{k\sigma\beta}\rangle }{(  \omega-\varepsilon_{-k})(  \omega+\varepsilon_{k})  -\delta_{\beta,S}\Delta^{2}},\notag\\
\label{A32}
\end{eqnarray}
\begin{eqnarray}
L(  \omega)  &&=\sum_{k,\beta}\tilde{V}_{\beta}^{2}\frac{-\langle c_{k\sigma\beta}^{\dag}c_{-k\bar{\sigma}\beta}^{\dag}\rangle (\omega-\varepsilon_{-k})  +\sigma\Delta\delta_{\beta,S}\langle c_{-k\bar{\sigma}\beta}^{\dag}c_{-k\bar{\sigma}\beta}\rangle }{(\omega-\varepsilon_{-k})  (  \omega+\varepsilon_{k})  -\delta_{\beta,S}\Delta^{2}}+\sum_{k,\beta}\tilde{V}_{\beta}^{2}\frac{-\langle c_{k\sigma\beta}^{\dag}c_{-k\bar{\sigma}\beta}^{\dag}\rangle (\omega-\varepsilon_{k})  +\sigma\Delta\delta_{\beta,S}\langle c_{k\sigma\beta}^{\dag}c_{k\sigma\beta}\rangle }{(  \omega-\varepsilon _{k})  (  \omega+\varepsilon_{-k})  -\delta_{\beta,S}\Delta^{2}}.\notag\\
\label{A33}
\end{eqnarray}
In this work, we focus on the large gap limit $\Delta\rightarrow\infty$ where above formulae can be greatly simplified while the physics we addressed are maintained. In this case, one has $\langle c_{k\sigma\beta}^{\dag}c_{k\sigma\beta}\rangle=\langle c_{-k\bar{\sigma}\beta}^{\dag}c_{-k\bar{\sigma}\beta}\rangle =\delta_{\beta,N}f_{N}(  \varepsilon_{k})  +\frac
{1}{2}\delta_{\beta,S}$ and $\langle c_{-k\bar{\sigma}\beta}c_{k\sigma\beta}\rangle =\langle c_{k\sigma\beta}^{\dag}c_{-k\bar{\sigma}\beta}^{\dag}\rangle =\delta_{\beta,S}\frac{\sigma}{2}$. Furthermore, in the wide-band
limit, the analytical results of the relevant summations can be obtained as $\sum_{k}\frac{\tilde{V}_{\beta}^{2}(  \omega+\varepsilon_{-k})  }{(  \omega-\varepsilon_{k})  (  \omega+\varepsilon
_{-k})  -\delta_{\beta,S}\Delta^{2}}=\sum_{k}\frac{\tilde{V}_{\beta}^{2}(  \omega-\varepsilon_{k})  }{(  \omega-\varepsilon_{k})(  \omega+\varepsilon_{-k})  -\delta_{\beta,S}\Delta^{2}}=\sum_{k}
\frac{\tilde{V}_{\beta}^{2}(  \omega-\varepsilon_{k}^{+})  }{(\omega-\varepsilon_{k}^{+})  (  \omega+\varepsilon_{-k}^{-})-\delta_{\beta,S}\Delta^{2}}=\sum_{k}\frac{\tilde{V}_{\beta}^{2}(\omega+\varepsilon_{-k}^{+})  }{(  \omega-\varepsilon_{k}^{-})  (\omega+\varepsilon_{-k}^{+})  -\delta_{\beta,S}\Delta^{2}}=-i\delta_{\beta,N}\frac{\tilde{\Gamma}_{N}}{2}$ and $\sum_{k}\frac{\tilde{V}_{\beta
}^{2}\sigma\Delta\delta_{\beta,S}}{(  \omega-\varepsilon_{k})  (\omega+\varepsilon_{-k})  -\delta_{\beta,S}\Delta^{2}}=\sum_{k}\frac{\tilde{V}_{\beta}^{2}\sigma\Delta\delta_{\beta,S}}{(  \omega-\varepsilon_{k}^{+})  (  \omega+\varepsilon_{-k}^{-})  -\delta_{\beta,S}\Delta^{2}}=\sum_{k}\frac{\tilde{V}_{\beta}^{2}\sigma\Delta\delta_{\beta,S}}{(  \omega-\varepsilon_{k}^{-})  (  \omega+\varepsilon_{-k}^{+})  -\delta_{\beta,S}\Delta^{2}}=-\sigma\delta_{\beta,S}\frac{\tilde{\Gamma}_{S}}{2}$. Thus, Eqs.\,(\ref{A6}), (\ref{A7}), (\ref{A26}) and (\ref{A27}) reduce to
\begin{equation}
(  \omega-\tilde{\varepsilon}_{d}+i\frac{\tilde{\Gamma}_{N}}{2})\langle \langle d_{\sigma};\hat{O}\rangle \rangle
=\langle [  d_{\sigma},\hat{O}]  _{+}\rangle-\sigma\frac{\tilde{\Gamma}_{S}}{2}\langle \langle d_{\bar{\sigma}}^{\dag};\hat{O}\rangle \rangle +U\langle \langle n_{d\bar{\sigma}}d_{\sigma};\hat{O}\rangle \rangle,\label{A34}
\end{equation}
\begin{equation}
(  \omega+\tilde{\varepsilon}_{d}+ i\frac{\tilde{\Gamma}_{N}}{2})\langle \langle d_{\bar{\sigma}}^{\dag};\hat{O}\rangle\rangle =\langle [  d_{\bar{\sigma}}^{\dag},\hat{O}]_{+}\rangle -\sigma\frac{\tilde{\Gamma}_{S}}{2}\langle \langle d_{\sigma};\hat{O}\rangle \rangle -U\langle \langle n_{d\sigma}d_{\bar{\sigma}}^{\dag};\hat{O}\rangle \rangle,\label{A35}
\end{equation}
\begin{eqnarray}
&&(\omega-\tilde{\varepsilon}_{d}-\tilde{U}+3i\frac{\tilde{\Gamma}_{N}}{2})\langle \langle n_{d\bar{\sigma}}d_{\sigma};\hat{O}\rangle \rangle\notag\\
&&=\langle [  n_{d\bar{\sigma}}d_{\sigma},\hat{O}]  _{+}\rangle -\left[\sum_{k}\frac{\tilde{V}_{N}^{2}f_{N}(  \varepsilon_{k})  }{\omega-\varepsilon_{k}
}+\sum_{k}\frac{\tilde{V}_{N}^{2}f_{N}(  \varepsilon_{k})  }{\omega+\varepsilon_{k}^{-}}\right]\langle \langle d_{\sigma};\hat{O}\rangle \rangle-\sigma\frac{\tilde{\Gamma}_{S}}{2}\langle \langle d_{\bar{\sigma}
}^{\dag};\hat{O}\rangle\rangle +\sigma\frac{\tilde{\Gamma}_{S}}{2}\langle \langle n_{d\sigma}d_{\bar{\sigma}}^{\dag};\hat{O}\rangle\rangle,\label{A36}
\end{eqnarray}
\begin{eqnarray}
&&(\omega+\tilde{\varepsilon}_{d}+\tilde{U}+3i\frac{\tilde{\Gamma}_{N}}{2})\langle \langle n_{d\sigma}d_{\bar{\sigma}}^{\dag};\hat{O}\rangle \rangle\notag\\
&&=\langle[  n_{d\sigma}d_{\bar{\sigma}}^{\dag},\hat{O}]  _{+}\rangle -\left[\sum_{k}\frac{\tilde{V}_{N}^{2}f_{N}(  \varepsilon_{k})  }{\omega+\varepsilon_{k}}+\sum_{k}\frac{\tilde{V}_{N}^{2}f_{N}(  \varepsilon_{k})  }{\omega-\varepsilon_{k}^{-}}\right] \langle\langle d_{\bar{\sigma}}^{\dag};\hat{O}\rangle \rangle -\sigma\frac{\tilde{\Gamma}_{S}}{2}\langle\langle d_{\sigma};\hat{O}\rangle \rangle +\sigma\frac
{\tilde{\Gamma}_{S}}{2}\langle \langle n_{d\bar{\sigma}}d_{\sigma};\hat{O}\rangle \rangle. \label{A37}
\end{eqnarray}
Replacing the operator $\hat{O}$ in Eqs.\,(\ref{A34})-(\ref{A37}) by $d_{\sigma}^{\dag}$ and $d_{\bar{\sigma}}$, respectively, we obtain following matrix equations
\begin{equation}
[  \mathbf{\tilde g}^{-1}_0(\omega)-\mathbf{\tilde\Sigma}_{0}(  \omega) ]  \left(
\begin{array}
[c]{cc}
\langle \langle d_{\sigma};d_{\sigma}^{\dag}\rangle\rangle  & \langle \langle d_{\sigma};d_{\bar{\sigma}}\rangle \rangle \\
\langle \langle d_{\bar{\sigma}}^{\dag};d_{\sigma}^{\dag}\rangle \rangle  & \langle \langle d_{\bar{\sigma}}^{\dag};d_{\bar{\sigma}}\rangle \rangle
\end{array}
\right)  =\mathbf I+\tilde{U}\left(
\begin{array}
[c]{cc}
\langle \langle n_{d\bar{\sigma}}d_{\sigma};d_{\sigma}^{\dag}\rangle\rangle  & \langle \langle n_{d\bar{\sigma}}d_{\sigma};d_{\bar{\sigma}}\rangle\rangle \\-\langle\langle n_{d\sigma}d_{\bar{\sigma}}^{\dag};d_{\sigma}
^{\dag}\rangle \rangle  & -\langle \langle n_{d\sigma}d_{\bar{\sigma}}^{\dag};d_{\bar{\sigma}}\rangle\rangle
\end{array}
\right),  \label{A38}
\end{equation}
and
\begin{equation}
\mathbf P^{-1}(  \omega)  \left(
\begin{array}
[c]{cc}
\langle \langle n_{d\bar{\sigma}}d_{\sigma};d_{\sigma}^{\dag}\rangle\rangle  & \langle \langle n_{d\bar{\sigma}}d_{\sigma};d_{\bar{\sigma}}\rangle\rangle \\
-\langle\langle n_{d\sigma}d_{\bar{\sigma}}^{\dag};d_{\sigma}^{\dag}\rangle \rangle  & -\langle \langle n_{d\sigma}d_{\bar{\sigma}}^{\dag};d_{\bar{\sigma}}\rangle\rangle
\end{array}
\right)  =\mathbf N+\mathbf Q(  \omega)  \left(
\begin{array}
[c]{cc}
\langle \langle d_{\sigma};d_{\sigma}^{\dag}\rangle \rangle  & \langle \langle d_{\sigma};d_{\bar{\sigma}}\rangle \rangle \\
\langle \langle d_{\bar{\sigma}}^{\dag};d_{\sigma}^{\dag}\rangle \rangle  & \langle \langle d_{\bar{\sigma}}^{\dag};d_{\bar{\sigma}}\rangle \rangle
\end{array}
\right),  \label{A39}
\end{equation}
where
\begin{eqnarray}
\mathbf {\tilde g}^{-1}_{0}(\omega)=\left(
\begin{array}
[c]{cc}
\omega-\tilde{\varepsilon}_{d} & 0\\
0 & \omega+\tilde{\varepsilon}_{d}
\end{array}
\right),
\end{eqnarray}
\begin{eqnarray}
\mathbf{\tilde{\Sigma}}_{0}(  \omega)  =\left(
\begin{array}
[c]{cc}
-i\frac{\tilde{\Gamma}_{N}}{2} & -\sigma\frac{\tilde{\Gamma}_{S}}{2}\\
-\sigma\frac{\tilde{\Gamma}_{S}}{2} & -i\frac{\tilde{\Gamma}_{N}}{2}
\end{array}
\right),
\end{eqnarray}
\begin{eqnarray}
\mathbf I=\left(
\begin{array}
[c]{cc}
1  & 0 \\
0  &1
\end{array}
\right),
\end{eqnarray}
\begin{eqnarray}
\mathbf N=\left(
\begin{array}
[c]{cc}
\langle n_{d\bar{\sigma}}\rangle  & \langle d_{\bar{\sigma}
}d_{\sigma}\rangle \\
\langle d_{\sigma}^{\dag}d_{\bar{\sigma}}^{\dag}\rangle  &
-\langle n_{d\sigma}\rangle
\end{array}
\right),
\end{eqnarray}
\begin{eqnarray}
\mathbf P^{-1}\left(  \omega\right)  =\left(
\begin{array}
[c]{cc}
\omega-\tilde{\varepsilon}_{d}-\tilde{U}+\frac{3i\tilde{\Gamma}_{N}}{2} &
\sigma\frac{\tilde{\Gamma}_{S}}{2}\\
\sigma\frac{\tilde{\Gamma}_{S}}{2} & \omega+\tilde{\varepsilon}_{d}+\tilde{U}
+\frac{3i\tilde{\Gamma}_{N}}{2}
\end{array}
\right),
\end{eqnarray}
and \begin{eqnarray}
\mathbf Q(\omega)  =\left(
\begin{array}
[c]{cc}
-\sum_{k}\frac{\tilde{V}_{N}^{2}f_{N}  (\varepsilon_{k})}{\omega-\varepsilon_{k}}-\sum_{k}\frac{\tilde{V}_{N}^{2}f_{N}(  \varepsilon_{k})  }{\omega+\varepsilon_{k}^{-}}  & -\sigma\frac{\tilde{\Gamma}_{S}}{2}\\
\sigma\frac{\tilde{\Gamma}_{S}}{2} & \sum_{k}\frac{\tilde{V}_{N}^{2}f_{N}(  \varepsilon_{k})  }{\omega+\varepsilon_{k}}+\sum_{k}\frac{\tilde{V}_{N}^{2}f_{N}(  \varepsilon_{k})  }{\omega-\varepsilon_{k}^{-}}
\end{array}
\right).
\end{eqnarray}
It is readily to check the relations that $\mathbf P_{22}(\omega)=-[\mathbf P_{11}(-\omega)]^*$, $\mathbf P_{21}(\omega)=\mathbf P_{12}(\omega)$, $\mathbf Q_{22}(\omega)=[\mathbf Q_{11}(-\omega)]^*$ and $\mathbf Q_{21}(\omega)=-\mathbf Q_{12}(\omega)$. Combining Eqs.\,(\ref{A38}) and (\ref{A39}) we finally arrive at
\begin{equation}
[  \mathbf{\tilde g}_{0}^{-1}(\omega)-\mathbf{\tilde{\Sigma}}_{0}(  \omega)  -\tilde{U} \mathbf{P}(\omega)  \mathbf{Q}(  \omega)  ] \mathbf{\tilde{G}}^{r}(\omega)=\mathbf{I}+\tilde U\mathbf{P}(  \omega)  \mathbf{N}. \label{A40}
\end{equation}
Note that the elements of matrix $\mathbf{N}$ should be calculated by the equations $\langle n_{d\sigma}\rangle=\langle n_{d\bar\sigma}\rangle=\int \frac{d\omega}{2\pi i}\mathbf{\tilde G}_{11}^{<}(\omega)$ and $\langle d_{\bar\sigma}d_{\sigma}\rangle=\langle d_{\sigma}^\dag d_{\bar\sigma}^\dag\rangle^*=\int \frac{d\omega}{2\pi i}\mathbf{\tilde G}_{12}^{<}(\omega)$, where the lesser and greater GFs are related to the retarded one through the Keldysh equation $\mathbf{{\tilde G}}^{<(>)}(\omega)=\mathbf{{\tilde G}}^{r}(\omega)\mathbf{{\tilde \Sigma}}^{<(>)}(\omega)\mathbf{{\tilde G}}^{a}(\omega)$ with $\mathbf{\tilde G}^a(\omega)=[\mathbf{\tilde G}^r(\omega)]^\dag$. In nonequilibrium situation, one of the proposed schemes to determine the interacting lesser (greater) self-energy $\mathbf{{\tilde\Sigma}}^{<(>)}(\omega)$ is the Ng's ansatz \cite{Ng1996} which has been widely employed in N-QD-N systems, as well as in N-QD-S cases \cite{Fazio1998,Sun2001}. However, the feasibility of this ansatz on the latter system is not such self-evident. Instead, we approximate $\mathbf{{\tilde\Sigma}}^{<(>)}(\omega)$ by their noninteracting counterparts $\mathbf{{\tilde\Sigma}}^{<(>)}_0(\omega)$ \cite{Sun2000}, which is considered to be reliable to capture our physical predictions at qualitative level. The formulae are thus closed, which can be self-consistently calculated to determine the GFs $\mathbf{\tilde G}^r(\omega)$, $\mathbf{\tilde G}^<(\omega)$, and $\mathbf{\tilde G}^>(\omega)$, and subsequently the current $I$, differential conductance $G\equiv dI/dV$, and LDOS $\rho(\omega)$, as formulated in the main text.

\begin{figure}[t]
\centering
\includegraphics[width=0.7\columnwidth]{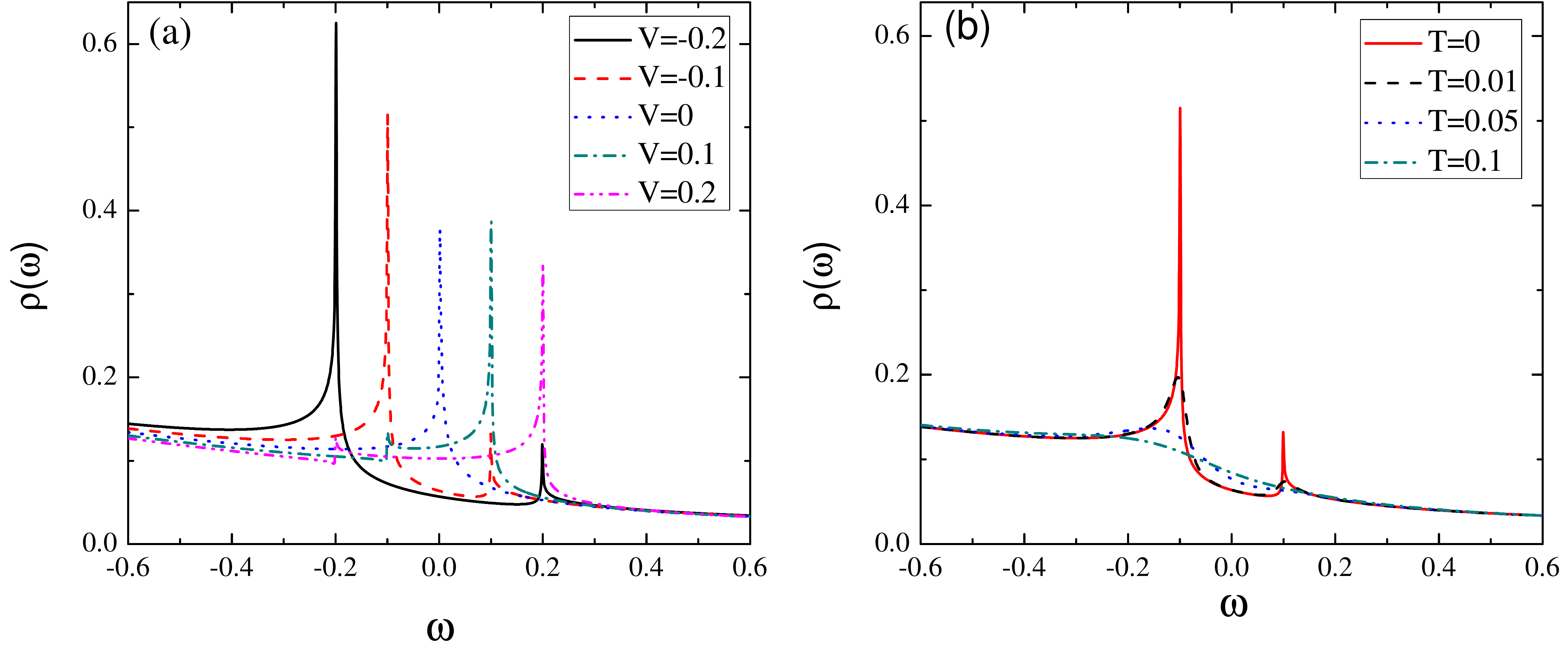}
\caption{(Color online) The Kondo resonances in the absence of EPI at different bias voltages with $T=0$ (a) and at different temperatures with $V=-0.1$ (b). Other parameters are $\varepsilon_d=-2.5$, $\Gamma_S=4$, and $U=10$.}\label{figs1}
\end{figure}

\subsection{Effects of bias voltage and temperature on the LDOS and the differential conductance}
In Fig.\,\ref{figs1}(a), we show the zero temperature Kondo resonances at different bias voltages in the absence of electron-phonon interaction (EPI). In equilibrium, a single Kondo resonance is pined at the Fermi level ($\mu_N=\mu_S=0$). As a finite bias voltage $V$ is applied to the normal lead, besides the main Kondo resonance aligned with $\mu_N=V$ another weak Kondo resonance develops at $\omega=-V$, whose physical originations are illustrated by the schematics $L_1$ and $R_1$, respectively, in Fig.\,2(c) in the main text. We note that these two Kondo resonances can not be reproduced within the truncation approximation made in Ref.\,[\onlinecite{Domanski2008}]. This is because the superconducting effects are not involved in the high-order GFs, since all the GFs in our Eq.\,(\ref{A9}) were discarded by the authors. However, the heights of the Kondo resonances in Fig.\,\ref{figs1}(a) are not suppressed by the decoherence effects induced by the bias as expected. This is a drawback of the EOM method in nonequilibrium problems \cite{Zimbovskaya2008}.

\begin{figure}[t]
\centering
\includegraphics[width=0.5\columnwidth]{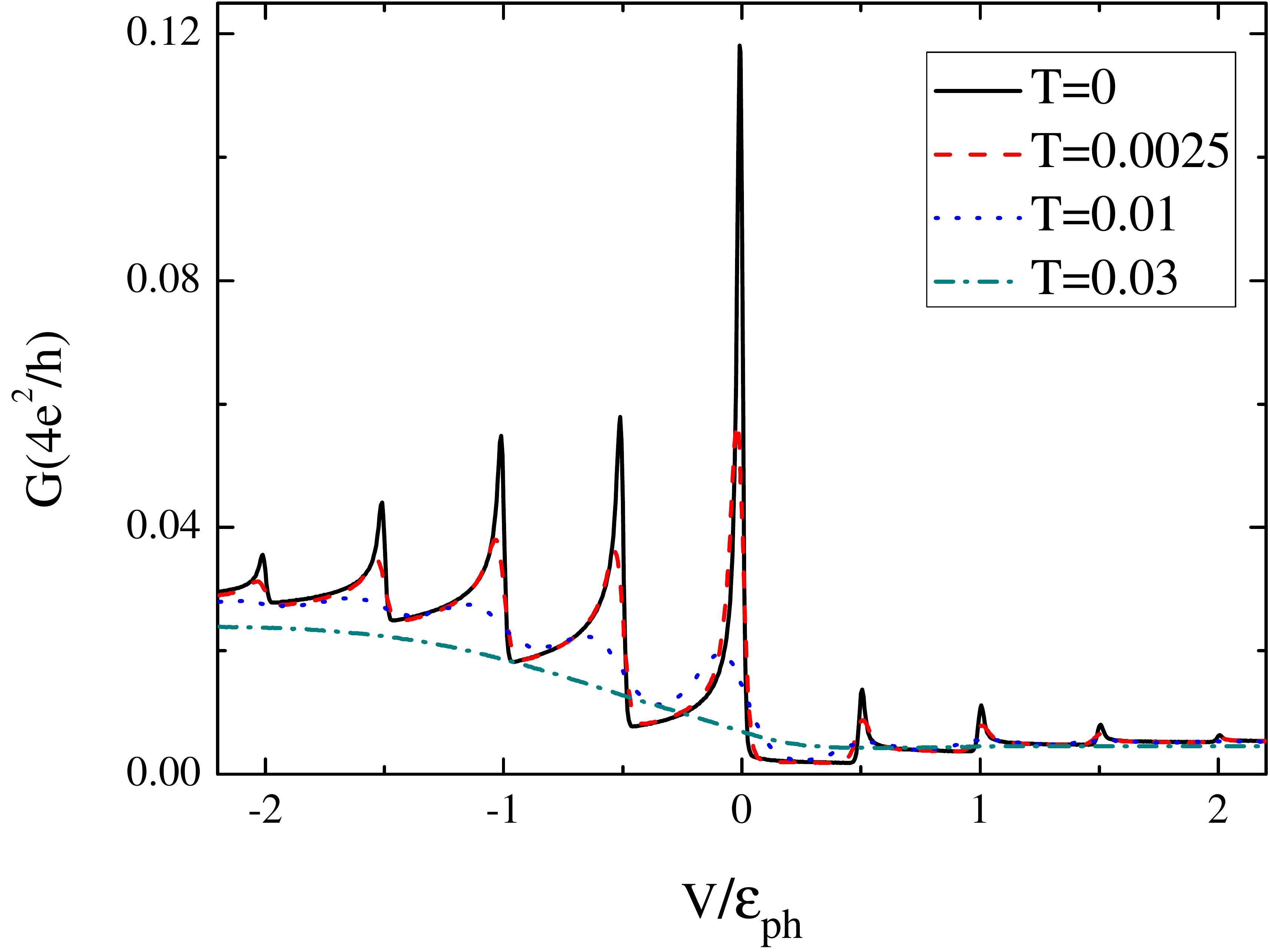}
\caption{(Color online) The differential conductance at different temperatures with finite EPI ($g=1.5$ and $\varepsilon_{ph}=0.1$). Other parameters are $\tilde\varepsilon_d=-2.5$, $\tilde\Gamma_S=4$, and $\tilde U=10$. Note that the curves here are not offset as in the Fig.\,2(a) in the main text.}\label{figs2}
\end{figure}

In this work, we do not attempt to estimate a Kondo temperature as the one in normal systems. This is because the true Kondo temperature of the full system is very difficult to define within the equation of motion approach or other theoretical methods. Of course, we could follow the usual procedure as some people did in the literature: use the normal Kondo temperature (pertaining to the reduced system with only normal leads) to characterize the Kondo effect in the full system which includes superconducting leads and even phonon modes. However, we do not think such a procedure would benefit a lot to the purpose of our present work, since the true Kondo temperature of the full system must be very different from the normal one. In order to obtain some qualitative insights, we present the effects of temperature on the Kondo resonances and the differential conductance. In Fig.\,\ref{figs1}(b), in agreement with the expected behavior for Kondo resonances, increasing the temperature suppresses the Kondo resonances intensively. In Fig.\,\ref{figs2}, as a supplement to the Fig.\,2(a) in the main text, we show the effect of temperature on the Kondo sidebands in the differential conductance. Clearly, as the temperature increases all the sidebands are broadened and suppressed simultaneously. Eventually, they disappear at all at high temperatures.

\section{Modified second-order perturbation theory in the Coulomb interaction}\label{SOPT}
As a complementary to the EOM method, we employ the SOPT in the Coulomb interaction proposed recently in Ref.\,[\onlinecite{Yamada2011}] to obtain the GFs, $\tilde{\mathbf{G}}_{11}^r(\omega)$, $\tilde{\mathbf{G}}_{11}^<(\omega)$, and $\tilde{\mathbf{G}}_{11}^>(\omega)$, for calculating the LDOS and the differential conductance. The formalism restricted to the spin degenerate dot level and the large gap limit is summarized as follows.

The retarded and advanced GFs can be obtained by the Dyson equation
\begin{equation}
\mathbf{\tilde G}^{r,a}(\omega)=\left[[\mathbf{\tilde G}_{0}^{r,a}(\omega)]^{-1}-\mathbf{\tilde\Sigma}_{\text{1st}}^{r,a}(\omega)-\mathbf{\tilde\Sigma}_{\text{2nd;mod}}^{r,a}(\omega)\right]^{-1}
\end{equation}
and the greater and lesser GFs can be obtained by the Keldysh equation
\begin{equation}
\mathbf{\tilde G}^{\gtrless}(\omega)=\mathbf{ \tilde G}^{r}(\omega)[\mathbf{\tilde\Sigma}_{\text{leads}}^{\gtrless}(\omega)+\mathbf{\tilde\Sigma}_{\text{2nd;mod}}^{\gtrless}(\omega)]\mathbf{\tilde G}^{a}(\omega),\label{eqs2}
\end{equation}
where
\begin{eqnarray}
[\mathbf{\tilde G}_{0}^{r,a}(\omega)]^{-1}=\left(
\begin{array}
[c]{cc}%
\omega-\tilde\varepsilon_{d}\pm i\tilde\Gamma_{N}/2 & \tilde\Gamma_{S}/2\\
\tilde\Gamma_{S}/2 & \omega+\tilde\varepsilon_{d}\pm i\tilde\Gamma_{N}/2
\end{array}
\right),\label{eqs3}
\end{eqnarray}
\begin{eqnarray}
\mathbf{\tilde\Sigma}_{\text{1st}}^{r,a}(\omega)=\tilde U\left(
\begin{array}
[c]{cc}%
\langle n_{d}\rangle  & \langle d_{\downarrow}d_{\uparrow}\rangle \\
\langle d_{\downarrow}d_{\uparrow}\rangle^*  &-\langle n_{d}\rangle
\end{array}\right),
\end{eqnarray}
\begin{eqnarray}
\mathbf{\tilde\Sigma}_{\text{leads}}^{<}(\omega)=i\tilde\Gamma_N\left(
\begin{array}
[c]{cc}%
f(\omega-e V)  & 0 \\
0  &f(\omega+e V)
\end{array}\right),
\end{eqnarray}
and
\begin{eqnarray}
\mathbf{\tilde\Sigma}_{\text{leads}}^{>}(\omega)=-i\tilde\Gamma_N\left(
\begin{array}
[c]{cc}%
f(-\omega+e V)  & 0 \\
0  &f(-\omega-e V)
\end{array}\right).
\end{eqnarray}
There is no first-order contribution to the lesser and greater self-energies because the Coulomb interaction in the QD takes place without delay. $\mathbf{\tilde\Sigma}_{\text{2nd;mod}}^{r,a}(\omega)=A\left[[\mathbf{\tilde\Sigma}_{\text{2nd}}^{r,a}(\omega)]^{-1}-B\right]^{-1}$ and $\mathbf{\tilde\Sigma}_{\text{2nd;mod}}^{\gtrless}(\omega)=\frac{1}{A}\mathbf{\tilde{\Sigma}}_\text{2nd;mod}^{r}(\omega)\left[\mathbf{\tilde\Sigma}_\text{2nd}^{r}(\omega)\right]  ^{-1}\mathbf{\tilde\Sigma}_\text{2nd}^{\lessgtr}(\omega)\left[\mathbf{\tilde\Sigma}_\text{2nd}^{a}(\omega)\right]^{-1}\mathbf{\tilde\Sigma}_\text{2nd;mod}^{a}(\omega)$ are the modified second-order selfenergies. The coefficient $A$ is determined so that $\mathbf{\tilde\Sigma}_{\text{2nd;mod}}^{r,a}(\omega)$ reproduce the leading behavior at high frequencies, and afterwards $B$ is determined to reproduce the exact result in the superconducting atomic limit ($\Delta\rightarrow\infty$, $\tilde\Gamma_N/\tilde U\rightarrow0$). Due to the interpolation of the second-order self-energies this method could provide results at arbitrary Coulomb interaction. The expressions of $A$ and $B$ are presented below. The second-order selfenergies $\mathbf{\tilde\Sigma}_{\text{2nd}}^{r,a}(\omega)$ and $\mathbf{\tilde\Sigma}_{\text{2nd}}^{\gtrless}(\omega)$ are given by
\begin{eqnarray}
\mathbf{\tilde\Sigma}_\text{2nd}^{\lessgtr}(\omega)=\tilde U^{2}\int\frac{d\omega_{1}}{2\pi}\tilde\Pi^{\lessgtr}(\omega+\omega_{1})\left(
\begin{array}
[c]{cc}
\mathbf{\tilde g}_{22}^{\gtrless}(  \omega_{1})   & -\mathbf{\tilde g}_{12}^{\gtrless}(\omega_{1})  \\
-\mathbf{\tilde g}_{21}^{\gtrless}(  \omega_{1})   & \mathbf{\tilde g}_{11}^{\gtrless}(\omega_{1})
\end{array}
\right)
\end{eqnarray}
and
\begin{equation}
\mathbf{\tilde\Sigma}_\text{2nd}^{r,a}(\omega)=\frac{i}{2\pi}\int\frac{d\omega_{1}}{\omega-\omega_{1}\pm i0^{+}}\left[\mathbf{\tilde\Sigma}_\text{2nd}^{>}(\omega_{1})-\mathbf{\tilde\Sigma}_\text{2nd}^{<}(\omega_{1})\right],
\end{equation}
where
\begin{equation}
\tilde\Pi^{\lessgtr}(\omega)=\int\frac{d\omega_{1}}{2\pi}\left[\mathbf{\tilde g}_{11}^{\lessgtr}(\omega_{1})\mathbf{\tilde g}_{22}^{\lessgtr}(\omega-\omega_{1})-\mathbf{\tilde g}_{12}^{\lessgtr}(\omega_{1})\mathbf{\tilde g}_{21}^{\lessgtr}(\omega-\omega_{1})\right],
\end{equation}
\begin{equation}
\mathbf{\tilde g}^{\gtrless}(\omega)=\mathbf{ \tilde g}^{r}(\omega)\mathbf{\tilde\Sigma}_{\text{leads}}^{\gtrless}(\omega)\mathbf{\tilde g}^{a}(\omega),
\end{equation}
and
\begin{eqnarray}
\mathbf{\tilde g}^{r,a}(\omega)=\left(
\begin{array}
[c]{cc}
\omega-\tilde\varepsilon_{d}-\tilde U\overline{\langle n_{d}\rangle }\pm i\tilde\Gamma_{N}/2 & \tilde\Gamma_{S}/2-\tilde U\overline{\langle d_{\downarrow}d_{\uparrow}\rangle }\\
\tilde\Gamma_{S}/2-\tilde U\overline{\langle d_{\downarrow}d_{\uparrow}\rangle}^* & \omega+\tilde\varepsilon_{d}+\tilde U\overline{\langle n_{d}\rangle }\pm i\tilde\Gamma_{N}/2
\end{array}
\right)^{-1}.\label{eq10}
\end{eqnarray}
$\overline{\langle n_{d}\rangle }$ and $\overline{\langle d_{\downarrow}d_{\uparrow}\rangle }$ in Eq.\,(\ref{eq10}) are two effective parameters which would be determined by the self-consistency conditions given below. From $\mathbf{\tilde g}^{<}(\omega)$ and $\mathbf{\tilde G}^{<}(\omega)$ we can obtain $\left\langle n_{d}\right\rangle_{0}=\int \frac{d\omega}{2\pi i} \mathbf{\tilde g}_{11}^{<}(\omega)$, $\langle d_{\downarrow}d_{\uparrow}\rangle_{0}=\int \frac{d\omega}{2\pi i} \mathbf{\tilde g}_{12}^{<}(\omega)$, $\langle n_{d}\rangle=\int \frac{d\omega}{2\pi i} \mathbf{\tilde G}_{11}^{<}(\omega)$, and $\left\langle d_{\downarrow}d_{\uparrow}\right\rangle=\int \frac{d\omega}{2\pi i} \mathbf{\tilde G}_{12}^{<}(\omega)$. Defining $\chi_{0}=\langle n_{d}\rangle _{0}(  1-\langle n_{d}\rangle _{0})-\vert \langle d_{\downarrow}d_{\uparrow}\rangle _{0}\vert ^{2}$ and $\chi=\langle n_{d}\rangle (1-\langle n_{d}\rangle)-\vert \langle d_{\downarrow}d_{\uparrow}\rangle \vert^{2}$, the coefficients $A$ and $B$ mentioned above can be obtained as $A={\chi}/{\chi_{0}}$ and
\begin{eqnarray}
B=\frac{1}{U\chi_{0}}\left(
\begin{array}
[c]{cc}
1-\langle n_{d}\rangle -\overline{\langle n_{d}\rangle } & -\langle d_{\downarrow}d_{\uparrow}\rangle -\overline{\langle d_{\downarrow}d_{\uparrow}\rangle }\\
-\langle d_{\downarrow}d_{\uparrow}\rangle^* -\overline{\langle d_{\downarrow}d_{\uparrow}\rangle}^* & -1+\langle n_{d}\rangle+\overline{\langle n_{d}\rangle }
\end{array}
\right).
\end{eqnarray}
In Ref.\,[\onlinecite{Yamada2011}], two self-consistent equations are given as
\begin{equation}
U\overline{\langle n_{d}\rangle }=U\langle n_{d}\rangle+\operatorname{Re}\left[\mathbf{\tilde\Sigma}_\text{2nd;mod}^{r}(\mu_N)\right]_{11},\label{eq12}
\end{equation}
and
\begin{equation}
U\overline{\langle d_{\downarrow}d_{\uparrow}\rangle }=U\langle d_{\downarrow}d_{\uparrow}\rangle+\left[\mathbf{\tilde\Sigma}_\text{2nd;mod}^{r}(\mu_S)
\right]_{12},\label{eq13}
\end{equation}
to calculate $\overline{\langle n_{d}\rangle}$, $\overline{\langle d_{\downarrow}d_{\uparrow}\rangle }$, $\langle n_{d}\rangle $, and $\langle d_{\downarrow}d_{\uparrow}\rangle $, simultaneously. In practical calculations, we find that Eqs.\,(\ref{eq12}) and (\ref{eq13}) are only good for quasi-equilibrium case. Instead, when the bias becomes larger, we use such self-consistent conditions as $\langle n_d\rangle_0=\langle n_d\rangle$ and $\langle d_{\downarrow}d_{\uparrow}\rangle_0=\langle d_{\downarrow}d_{\uparrow}\rangle$ for calculations, which have been used in previous work \cite{Yeyati1993}. After obtaining the GFs, $\tilde{\mathbf{G}}_{11}^r(\omega)$, $\tilde{\mathbf{G}}_{11}^<(\omega)$, and $\tilde{\mathbf{G}}_{11}^>(\omega)$, one can get the full GFs, ${\mathbf{G}}_{11}^r(\omega)$, ${\mathbf{G}}_{11}^<(\omega)$, and ${\mathbf{G}}_{11}^>(\omega)$, through the relations
\begin{equation}
\mathbf{G}_{11}^{r}(\omega)=\sum_{n=-\infty}^{\infty}L_{n}[\mathbf{\tilde{G}}_{11}^{r}(\omega-n\varepsilon_{ph})
+\frac{1}{2}\mathbf{\tilde{G}}_{11}^{<}(\omega-n\varepsilon_{ph})-\frac{1}{2}\mathbf{\tilde{G}}_{11}^{<}(\omega+ n\varepsilon_{ph})]
\end{equation}
and
\begin{equation}\mathbf{G}_{11}^{<(>)}(\omega)=\sum_{n=-\infty}^{\infty}L_{n}\mathbf{\tilde{G}}_{11}^{<(>)}(\omega\pm n\varepsilon_{ph})
\end{equation}
mentioned in the main text. Subsequently, the LDOS and the differential conductance can be directly calculated.

\begin{figure}[t]
\centering
\includegraphics[width=0.7\columnwidth]{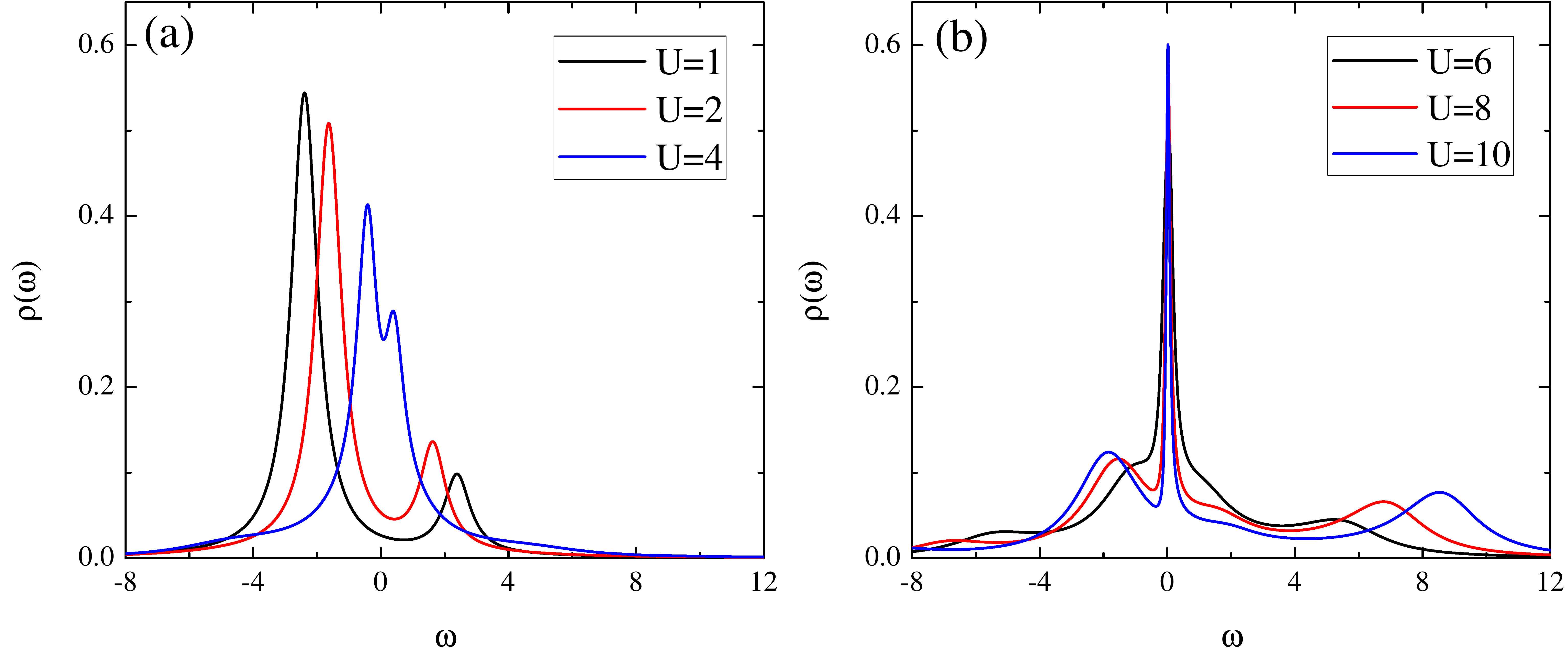}
\includegraphics[width=0.7\columnwidth]{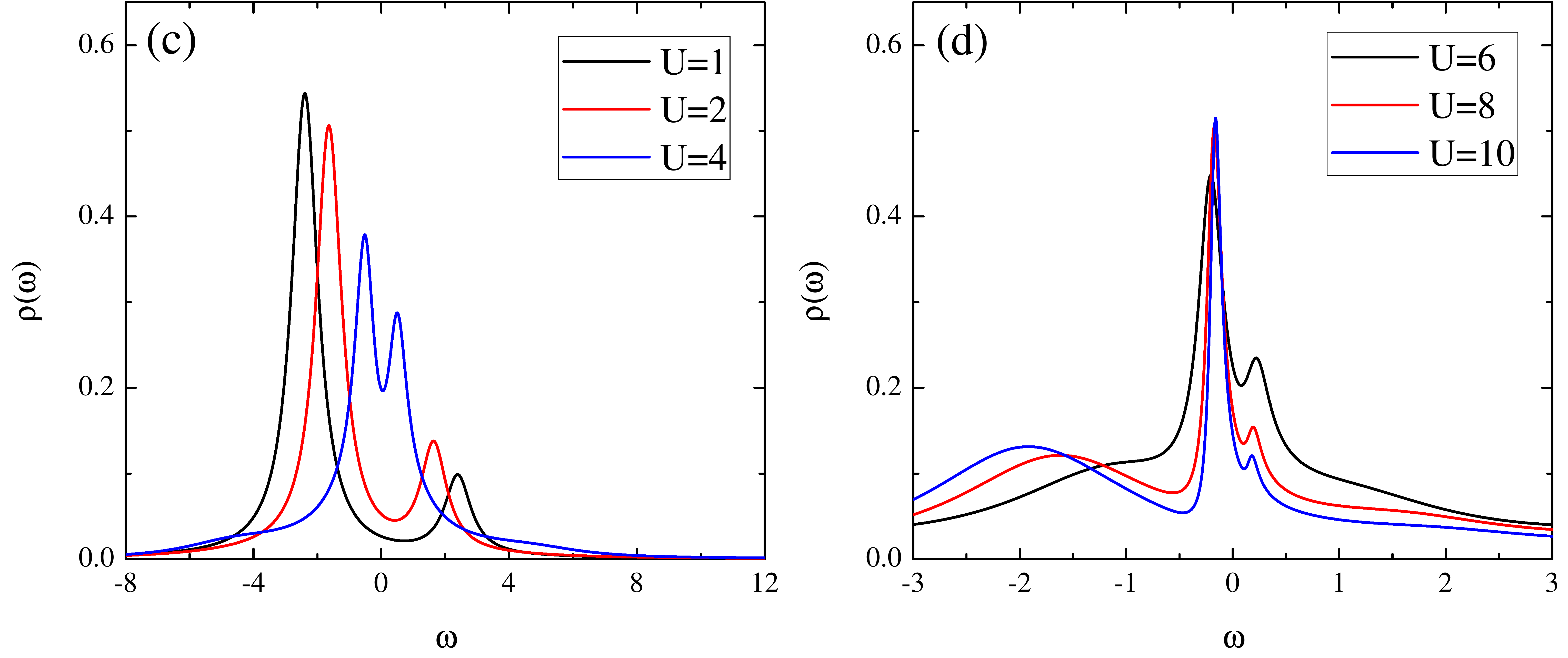}
\caption{(Color online) Upper and lower panels correspond to the equilibrium ($V=0$) and nonequilibrium ($V=-0.2$) LDOS, respectively, at different $U$ for fixed dot level $\varepsilon_d=-2.5$. Other parameters are $\Gamma_S=4$, $T=0.002$, and $g=0$. $\Gamma_N$ is taken as the energy unit.}\label{figs3}
\end{figure}

\begin{figure}[h]
\centering
\includegraphics[width=0.7\columnwidth]{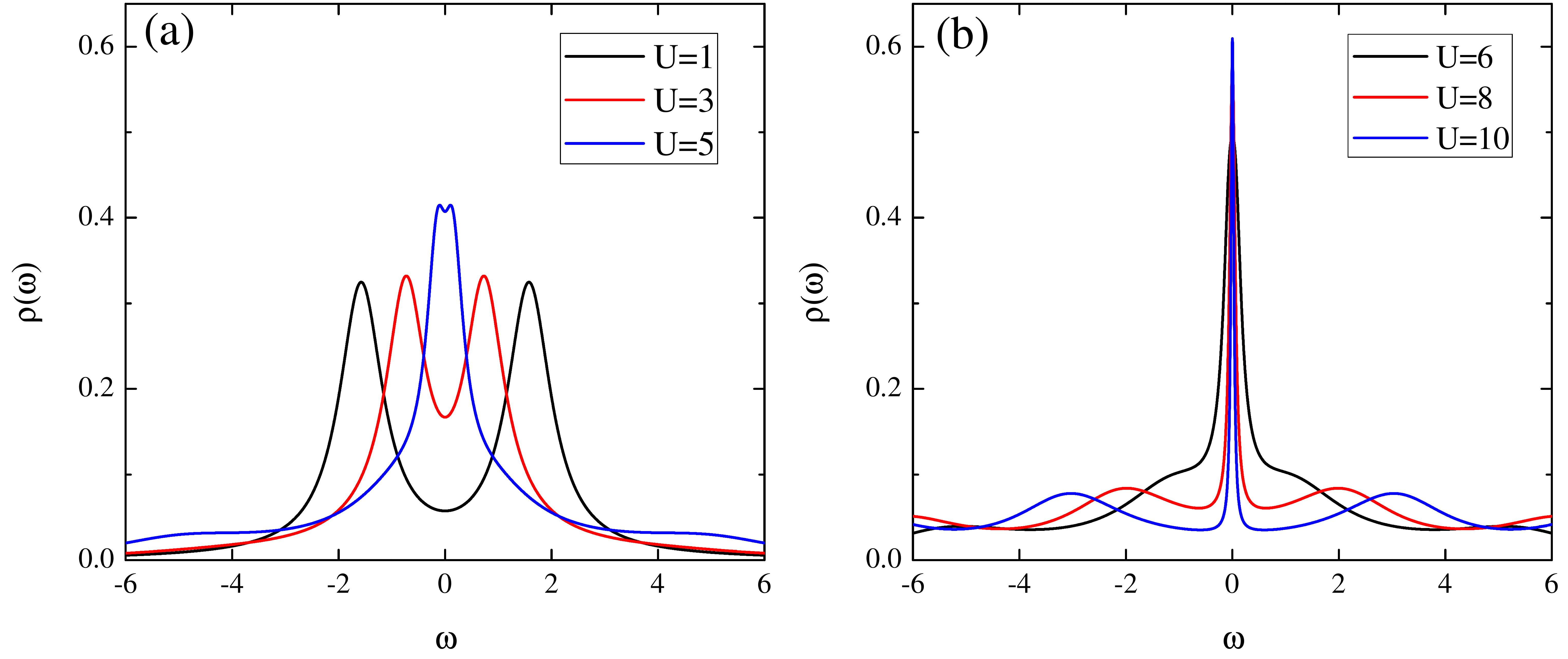}
\includegraphics[width=0.7\columnwidth]{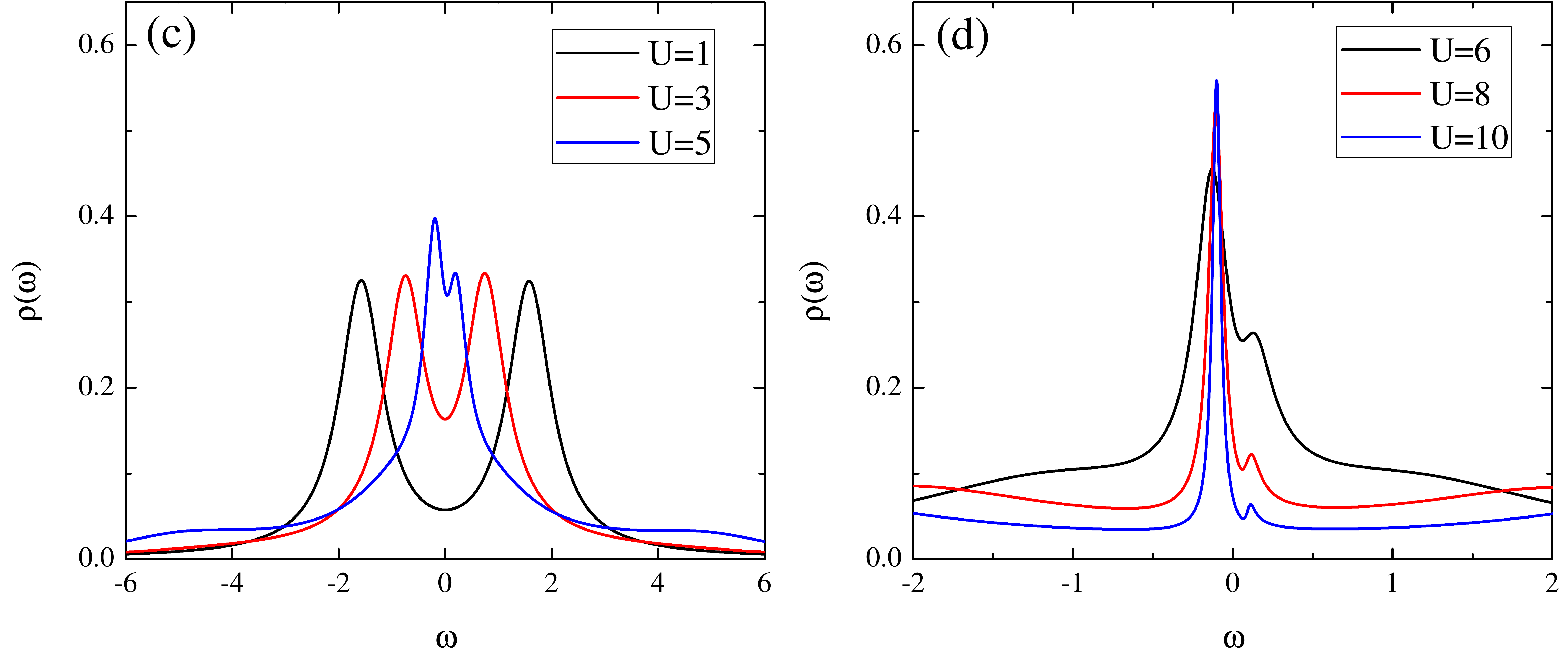}
\caption{(Color online) Upper and lower panels correspond to the equilibrium ($V=0$) and nonequilibrium ($V=-0.1$) LDOS, respectively, at different $U$ for fixed ratio $\varepsilon_d/U=-1/2$. Other parameters are $\Gamma_S=4$, $T=0.002$, and $g=0$.}\label{figs4}
\end{figure}

We first show the equilibrium and nonequilibrium LDOS in the absence of EPI. In Fig.\,\ref{figs3}, the effects of Coulomb interaction $U$ and bias voltage $V$ on the LDOS are displayed under a fixed dot level. At equilibrium, for a small $U$ there exits two Andreev bound states (ABSs) locating symmetrically on the two sides of the Fermi level, which move towards the Fermi level as $U$ increases [Fig.\,\ref{figs3}(a)]. In this case, the intradot pairing induced by the superconducting proximity effect dominates and the ground state of the system is the BCS singlet \cite{Yamada2011,Zitko2015}. For a larger $U$, the two ABSs even get across the Fermi level and since then a pronounced Kondo resonance develops at the Fermi level [Fig.\,\ref{figs3}(b)], indicating that the ground state of the system is the Kondo singlet \cite{Yamada2011,Zitko2015}. The crossover from BCS singlet to Kondo singlet can also be induced by adjusting the N-QD and QD-S tunnel couplings \cite{Domanski2016}. When a small bias $V$ is applied to the normal lead, the entire LDOS in the BCS singlet phase is insensitive to the bias [Fig.\,\ref{figs3}(c)], on the contrary, one can see in the Kondo singlet phase that the position of the Kondo resonance is shifted to $\omega=V$ and meanwhile another weak Kondo resonance develops at $\omega=-V$[Fig.\,\ref{figs3}(d)]. Similar behaviors persist in Fig.\,\ref{figs4} where both the dot level and the Coulomb interaction are changed while their ratio is kept as $\varepsilon_d/U=-1/2$.

\begin{figure}[b]
\centering
\includegraphics[width=0.7\columnwidth]{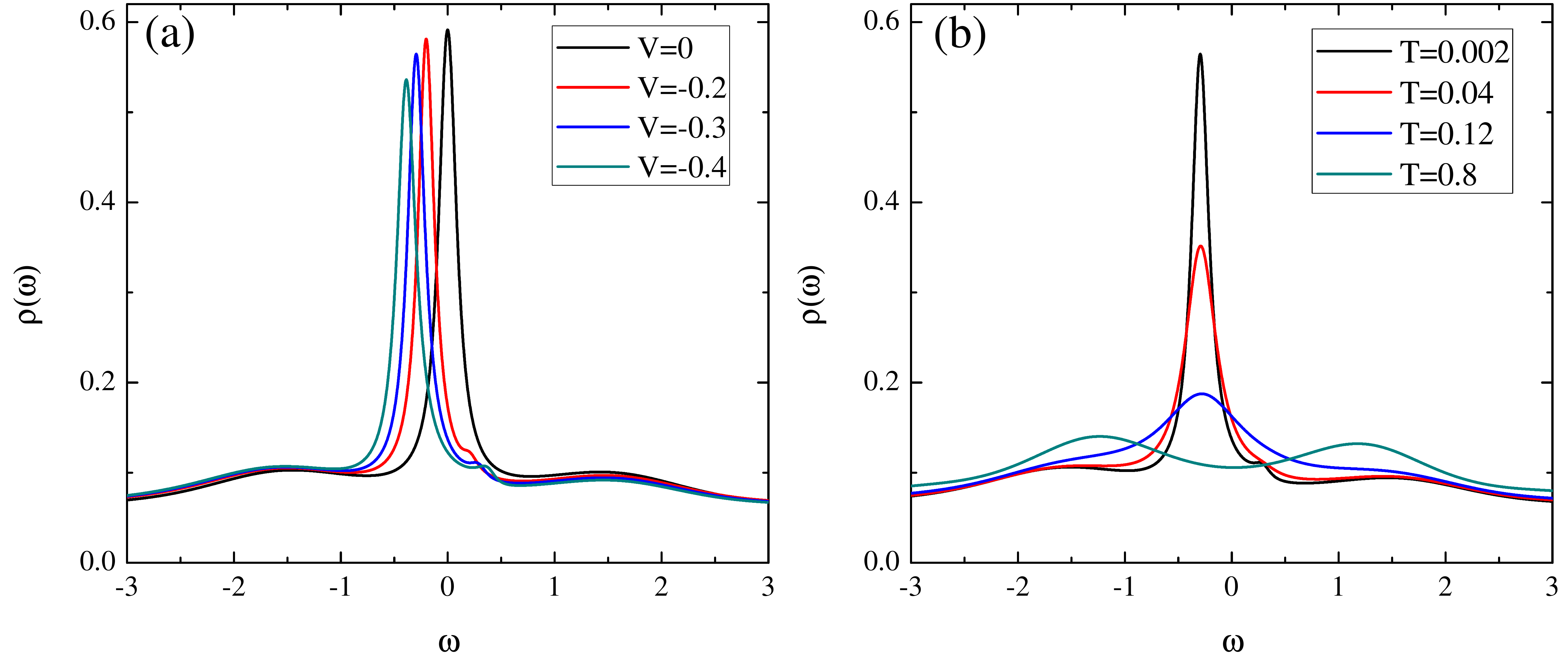}
\caption{(Color online) Effects of (a) bias (at $T=0.002$) and (b) temperature (at $V=-0.3$) on the LDOS. Other parameters are $U=5$, $\varepsilon_d=-2.5$, $\Gamma_S=2$, and $g=0$.}\label{figs5}
\end{figure}

As shown in Fig.\,\ref{figs5}, the sharp peaks appearing in Figs.\,\ref{figs3}(b), \ref{figs3}(d), \ref{figs4}(b), and \ref{figs4}(d) are identified as Kondo resonances. This is because they are intensively suppressed as the bias and the temperature increase, while the high-energy resonant peaks are almost unchanged.

\begin{figure}[t]
\centering
\includegraphics[width=0.7\columnwidth]{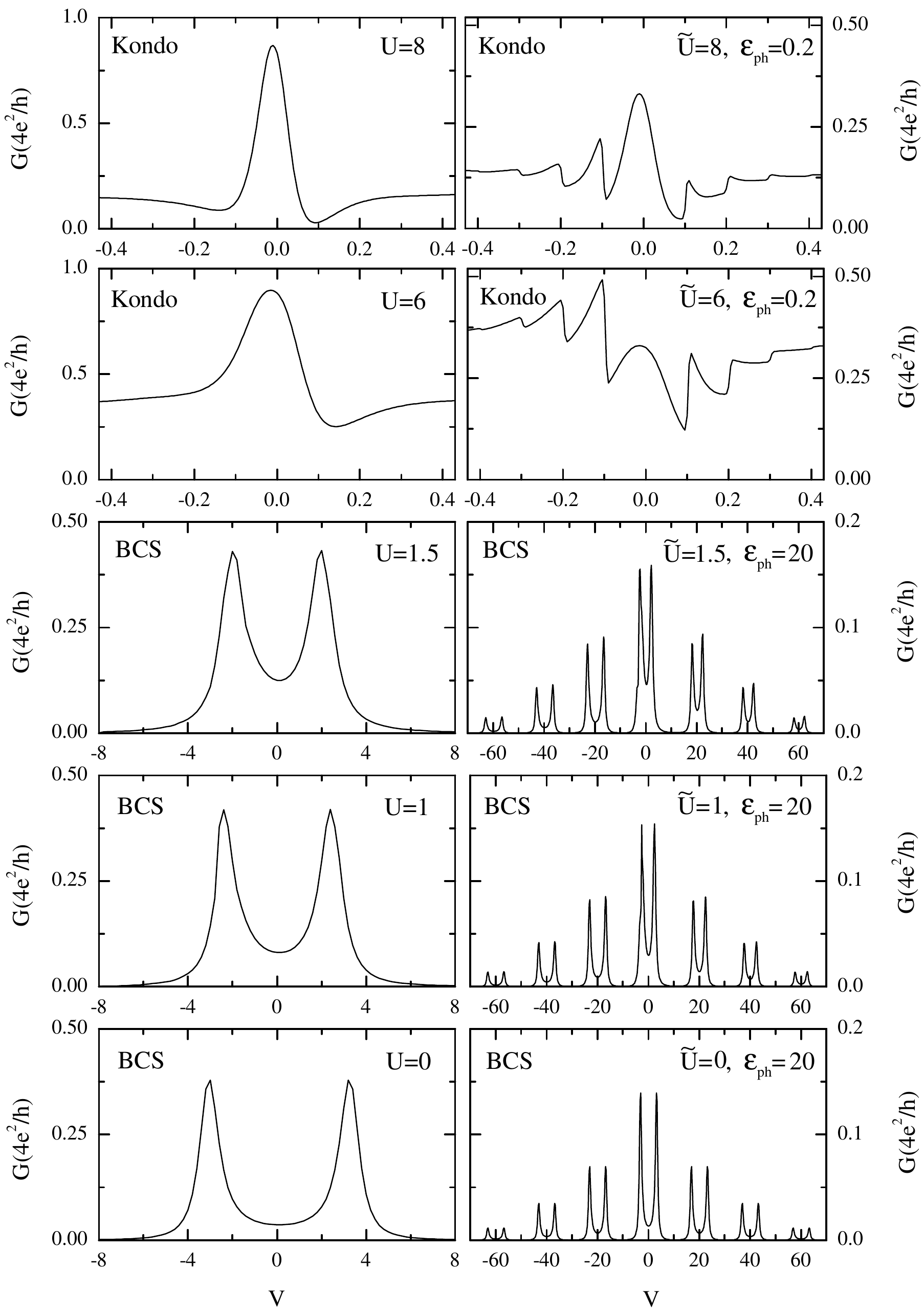}
\caption{(Color online) Differential conductance without (left panel) and with (right panel) the EPI at different $U$ for fixed dot level $\varepsilon_d=-2.5$. Here $\Gamma_S=4$ and $T=0.002$. In the presence of EPI ($g=1$) we set the renormalized parameters $\tilde U$, $\tilde \varepsilon_d$, and $\tilde \Gamma_{S(N)}$ as the same as $U$, $\varepsilon_d$, and $\Gamma_{S(N)}$, respectively. }\label{figs6}
\end{figure}

\begin{figure}[t]
\centering
\includegraphics[width=0.7\columnwidth]{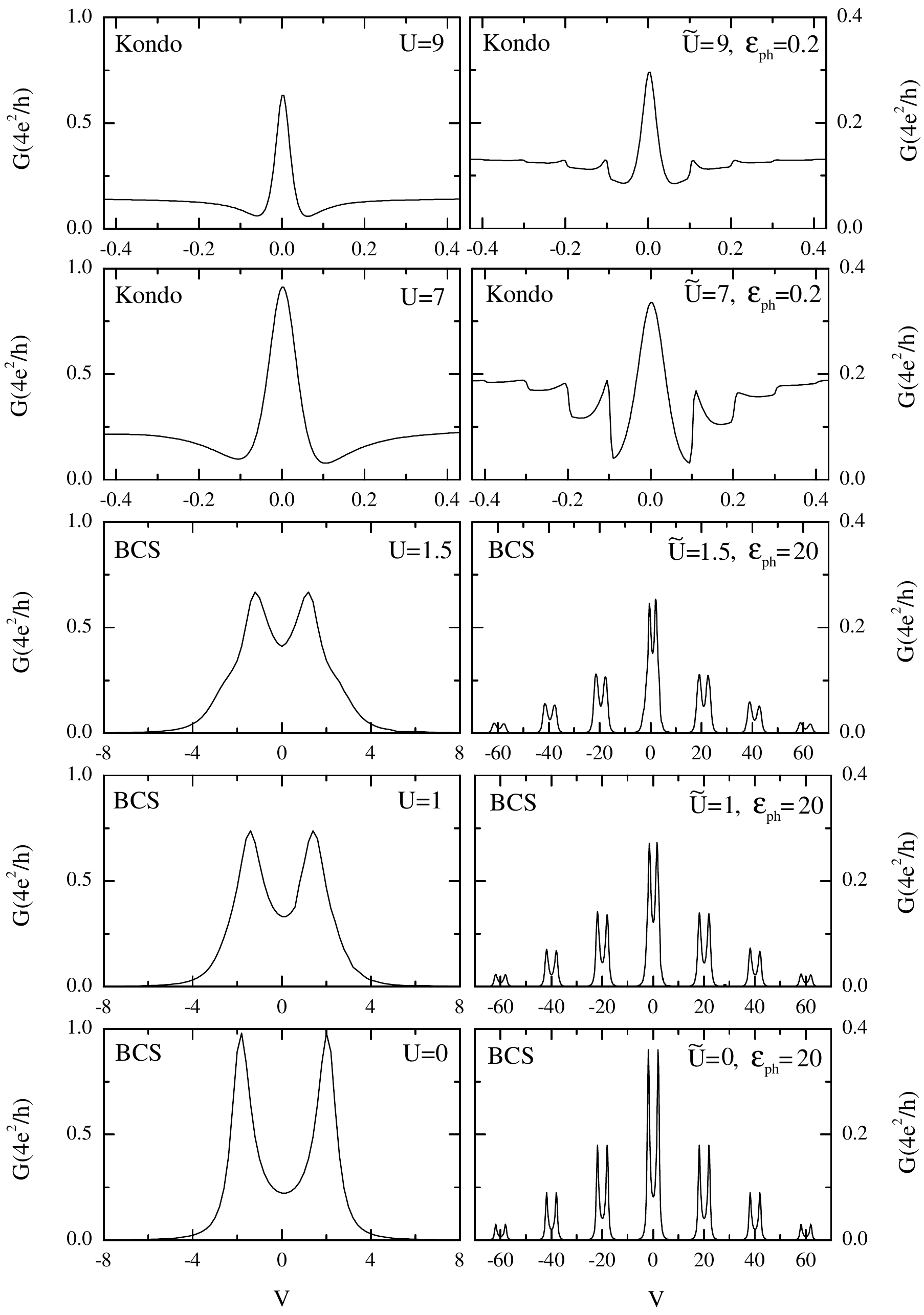}
\caption{(Color online) Differential conductance without (left panel) and with (right panel) the EPI at different $U$ for fixed ratio $\varepsilon_d/U=-1/2$. Other parameters are the same as in Fig.\,\ref{figs6}.}\label{figs7}
\end{figure}

Finally, the differential conductance without and with the EPI at large and small Coulomb interaction are displayed in Figs.\,\ref{figs6} and \ref{figs7} for fixed $\varepsilon_d$ and fixed ratio of $\varepsilon_d/U=-1/2$, respectively. It is clearly shown that the Kondo sidebands in the differential conductance induced by the EPI in the Kondo singlet phase are separated by half a phonon energy. In the BCS singlet phase, the relevant energy scale is no longer the width of the Kondo resonance. In this case, we set the phonon energy to be large enough (determined by the widths of ABSs) in order to clarify the sidebands in the differential conductance. As we can see, there are two interleaved sets of phonon sidebands in the differential conductance, each of which is separated by one phonon energy, as long as the system is still in the BCS singlet phase. Furthermore, in comparison with the $\tilde U=0$ case, a weak but nonzero $\tilde U$ will only affect the height and distance between the two sets of sidebands, while the qualitative behavior is unchanged. Note that the differential conductance at intermediate Coulomb interaction which corresponds to the BCS-Kondo crossover regime is not displayed. This is because stable and convergent self-consistent solutions can not be obtained in the crossover regime. This problem of the modified second-order perturbation theory also exists in the numerical calculations performed in Ref.\,[\onlinecite{Yamada2011}]. It is apparent that the phonon sideband features obtained by the SOPT here agree with those obtained by the  EOM method in the main text. This confirms the correctness of our main finding in this work, i.e., the phonon sidebands of the Kondo resonance being separated by half a phonon energy in the differential conductance.